\documentclass[11pt,english]{article}
\usepackage[T1]{fontenc}
\usepackage[latin9]{inputenc}
\usepackage{verbatim}
\usepackage{mathtools}
\usepackage{amsmath}
\usepackage{amssymb}
\usepackage{stackrel}
\usepackage{graphicx}
\usepackage[numbers]{natbib}

\makeatletter

\pdfoutput=1
\usepackage[T1]{fontenc}
\usepackage[latin9]{inputenc}
\usepackage{color}
\usepackage{float}
\usepackage{graphicx}
\usepackage{graphics}
\usepackage{esint}
\usepackage{enumerate}

\makeatletter

\usepackage{latexsym}
\usepackage{epsfig}
\usepackage{graphicx}
\usepackage[colorlinks,linkcolor=blue]{hyperref}
\allowdisplaybreaks 

\usepackage{babel}
\usepackage{tcolorbox}

\usepackage{tikz}

\makeatother

\usepackage{babel}
\begin{document}
\vspace*{10.0ex}
\centerline{\Large \bf From off-shell strings to entropy flows}

\vspace*{10.0ex}
\centerline{\large  Houwen Wu and Shuxuan Ying}
\vspace*{7.0ex}
\vspace*{2.0ex}

\centerline{\large \it College of Physics}
\centerline{\large \it Sichuan University}
\centerline{\large \it Chengdu, 610065, China} \vspace*{1.0ex}

\vspace*{3.0ex}
\centerline{\large \it Department of Physics}
\centerline{\large \it Chongqing University}
\centerline{\large \it Chongqing, 401331, China} \vspace*{1.0ex}
\vspace*{3.0ex}

\centerline{iverwu@scu.edu.cn, ysxuan@cqu.edu.cn}
\vspace*{8.0ex}
\centerline{\bf Abstract} \bigskip \smallskip
 
Bit threads express holographic entropy as the maximal flux of a divergenceless vector field subject to a local capacity bound, but do not identify a physical carrier of that flux. In our previous work, we identified the projected string charge density with a bit-thread flow, reproduced the entropies of BTZ and ten-dimensional three-charge black holes, and showed that the maximal entropy flux is preserved under S- and T-dualities. However, the norm bound was imported from the bit-thread prescription, while the string-current configurations were prescribed along known bit-thread lines. In this paper, we give a stringy foundation for the local capacity and construct dynamical probe-string representatives of maximal flow. At genus zero and two-derivative order, the conical variation of the target-space action derived from the off-shell sphere fixes the string-frame entropy density. The gravitational replica construction and weighted flow-cut duality then yield the string-frame norm bound. We next find regular stationary F1 solutions in an NS-NS supported BTZ black hole whose spatial parts are logarithmic spirals. A uniform ensemble produces a conserved, constant-norm projected string charge density and realizes a maximal flow whose flux equals the BTZ entropy. We finally extend the construction to BTZ backgrounds with mixed NS-NS and RR flux. Varying the flux fraction changes the stationary F1 configurations at fixed metric and dilaton, while the corresponding conserved flows reproduce the same leading BTZ entropy. Thus, both the norm bound and its physical maximal-flow representative arise from string theory, with stationary strings providing physical channels for the entropic flux represented geometrically by bit threads.

\vfill 
\eject
\baselineskip=16pt
\vspace*{10.0ex}
\tableofcontents

\section{Introduction}

\label{sec:introduction}

The Ryu--Takayanagi (RT) formula translates quantum entanglement
into bulk geometry \cite{Ryu:2006bv,Ryu:2006ef}. On a static bulk
slice, its bit-thread formulation replaces the RT minimal surface
by the maximum flux of a divergenceless vector field \cite{Freedman:2016zud,Headrick:2017ucz},
\begin{equation}
S\left(A\right)=\max_{v}\int_{A}v^{i}n_{i}dA_{E},\qquad\nabla_{i}^{E}v^{i}=0,\qquad|v|_{E}\leq\frac{1}{4G_{N}},\label{eq:intro-bitthreads}
\end{equation}
where the metric, norm, and area element are written in the Einstein
frame. The two constraints in this configuration have distinct roles.
The divergence condition makes the flux invariant under deformations
among homologous cuts, while the norm bound sets the largest entropy
flux that can pass through a unit area. Their combination turns the
RT surface into the bottleneck of a maximal flow and has led to direct
geometric proofs of holographic entropy inequalities. Related developments
include geometric constructions, higher-curvature extensions, thermal
bit threads, and recent string-diagrammatic and Lorentzian thread
formulations \cite{Agon:2018lwq,Harper:2018sdd,Caggioli:2024uza,Lin:2026ehl,Caceres:2025ypk}.
The vector $v^{i}$, however, enters the usual construction as the
dual variable of a convex optimization problem. Its integral curves
may be drawn as ``threads,'' but the flow formalism alone does not
identify a spacetime charge whose physical carriers follow those curves.

String theory supplies a natural candidate. A fundamental string is
electrically charged under the Kalb--Ramond two-form. In the three-dimensional
setting, varying its worldsheet coupling to $B_{\mathrm{ab}}$ and
dividing out the overall F1 tension gives the antisymmetric spacetime
current \cite{Dabholkar:1990yf,Sen:1992yt,Duff:1994an,Lu:2025awh}
\begin{equation}
j^{\mathrm{ab}}=\frac{1}{\sqrt{-g}}\int d^{2}\sigma\left(\frac{\partial X^{\mathrm{a}}}{\partial\tau}\frac{\partial X^{\mathrm{b}}}{\partial\sigma}-\frac{\partial X^{\mathrm{b}}}{\partial\tau}\frac{\partial X^{\mathrm{a}}}{\partial\sigma}\right)\delta^{\left(3\right)}\left(x-X\left(\tau,\sigma\right)\right).\label{eq:intro-string-current}
\end{equation}
For a closed worldsheet, or locally away from open-string endpoints,
this current is conserved. On a static slice with induced string-frame
metric $h_{ij}$ and ADM lapse $N_{\mathrm{ADM}}$, the components
with one temporal index define the projected string charge density
\begin{equation}
q^{i}=N_{\mathrm{ADM}}j^{0i},\qquad\partial_{i}\left(\sqrt{h}q^{i}\right)=0.\label{eq:intro-projected-current}
\end{equation}
The vector $q^{i}$ is tangent to the spatial string curve, and its
flux measures oriented Kalb--Ramond charge. After coarse-graining
a family of strings, it therefore has the conservation law and line
structure required of a bit-thread flow.

\begin{figure}[h]
\centering \includegraphics[scale=0.35]{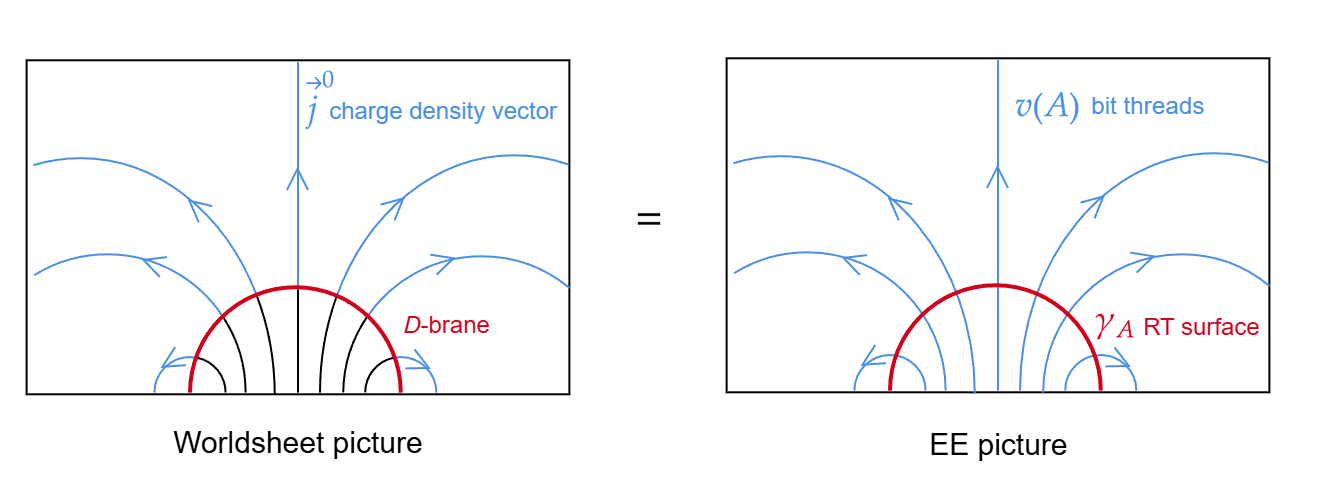} \caption{\label{fig:bit}The string charge density/bit threads correspondence.
The projected Kalb--Ramond charge density is tangent to the strings
and supplies conserved flow lines. The metric determines the transverse
geometry and the entropy bottleneck. In the original construction,
both the string arrangement and its entropy normalization were chosen
to reproduce a known bit-thread flow. The red entangling brane/D-brane
shown here imposes an auxiliary endpoint condition associated with
the RT surface $\gamma_{A}$.}
\end{figure}

Motivated by these properties, in ref. \cite{Wu:2025qwc} we proposed
the string charge density/bit threads correspondence 
\begin{equation}
v^{i}=C_{\mathrm{geom}}q^{i}.\label{eq:intro-original-correspondence}
\end{equation}
The spacetime metric determines the transverse geometry and the minimal
surface that forms the bottleneck, while the Kalb--Ramond current
supplies conserved lines passing through it. Schematically, 
\begin{equation}
g_{\mathrm{ab}}\quad\longleftrightarrow\quad\text{RT geometry},\qquad B_{\mathrm{ab}}\quad\longleftrightarrow\quad\text{string charge flow}\quad\longleftrightarrow\quad\text{bit threads}.\label{eq:intro-schematic-correspondence}
\end{equation}
In the original BTZ construction, $C_{\mathrm{geom}}$ converted one
unit of coarse-grained string charge $q^{i}$ into entropy flux $v^{i}$.
Requiring the flow to saturate the bottleneck capacity reproduced
the black hole entropy and gave a physical charge-current interpretation
to the otherwise auxiliary bit-thread lines. Fig. (\ref{fig:bit})
illustrates this proposal.

This correspondence is also naturally related to the Susskind--Uglum
worldsheet picture of gravitational entropy \cite{Susskind:1994sm}.
In that picture, black hole entropy is described by closed string
sphere diagrams whose target space embeddings intersect the horizon.
A closed string that crosses the horizon twice appears to an exterior
Rindler observer as an open segment with endpoints on the horizon.
Related entangling brane constructions make this open-string endpoint
picture precise in two-dimensional string theory \cite{Donnelly:2016jet}.
The closed string description associates entropy with the response
of the sphere amplitude to a conical deformation, while the open string
picture suggests physical lines attached to the entangling surface.
The string charge density construction gives these lines a spacetime
current and turns their spatial projection into a conserved entropy-flow
candidate.

Ahmadain and Wall placed the closed string part of this picture on
a precise off-shell footing \cite{Ahmadain:2022tew,Ahmadain:2022eso},
and it was subsequently extended with Frenkel \cite{Ahmadain:2024hdp}.
Using Tseytlin's sphere prescription, they derived the leading closed
string effective action from a regulated genus-zero worldsheet functional
and evaluated it on conical target space geometries. The cone is off
shell because its tip carries a localized worldsheet beta function,
or equivalently a target space graviton tadpole. Differentiating the
sphere-derived action with respect to the angular period gives the
Bekenstein--Hawking entropy at leading order in $\alpha^{\prime}$.
This result fixes the gravitational entropy associated with a prescribed
horizon directly from the closed string sphere. Related string and
worldsheet based analyses of entanglement and black hole entropy include
refs. \cite{He:2014gva,Brustein:2022wiq,Halder:2023adw,Mori:2025qoh,Ahmadain:2025pox}.

Our previous work supplied the complementary conserved-current description.
The correspondence was first formulated in an effective three-dimensional
setting and was subsequently extended to ten-dimensional three-charge
black holes. In the Type IIB S-duality analysis of the F1--NS5--P
and D1--D5--P systems \cite{Wu:2026qha}, the F1 current is mapped
to a D1 current. The metric, dilaton, local current, and pointwise
flow norm change, while the integrated maximal entropy flux is preserved.
Under T-duality \cite{Wu:2026kci}, a radial F1 remains an F1 along
the NS--NS duality chain, whereas a D1 transverse to the duality
circle becomes a D2 wrapped on the dual circle. Contracting the projected
D2 current with the normalized one-form along that circle gives an
effective conserved vector flow. The local representatives again change,
but the dilaton-weighted transverse capacity and the maximal entropy
flux remain invariant. These duality checks show that distinct string
and brane currents can represent the same physical entropy flux.

The original BTZ construction provided a conserved-current representation
of entropy flow, while taking the source embeddings and the local
entropy capacity as inputs. Specifically, as illustrated in fig. (\ref{fig:bit}),
it looks as if we artificially place parallel strings along the known
bit-thread lines, which would naturally give the desired flow. Current
conservation ensures that such prescribed strings define a divergenceless
vector field, but it does not show that their embeddings solve the
F1 equations in the specified background. Moreover, the conversion
coefficient in Eq. (\ref{eq:intro-original-correspondence}) was fixed
by imposing 
\begin{equation}
|C_{\mathrm{geom}}q|\leq\frac{e^{-2\phi\left(x\right)}}{4G_{N}^{\left(D\right)}},\label{eq:intro-old-bound}
\end{equation}
obtained from bit-thread formulation and demanding saturation at the
horizon. The entropy normalization was therefore provided by the geometric
capacity constraint, rather than independently determined within the
string construction. The present work derives this capacity from the
off-shell sphere functional, together with the gravitational replica
construction and flow--cut duality.

Therefore, the aim of this paper is to resolve these two issues at
genus zero and two-derivative order through the following steps: 

\includegraphics[scale=0.3]{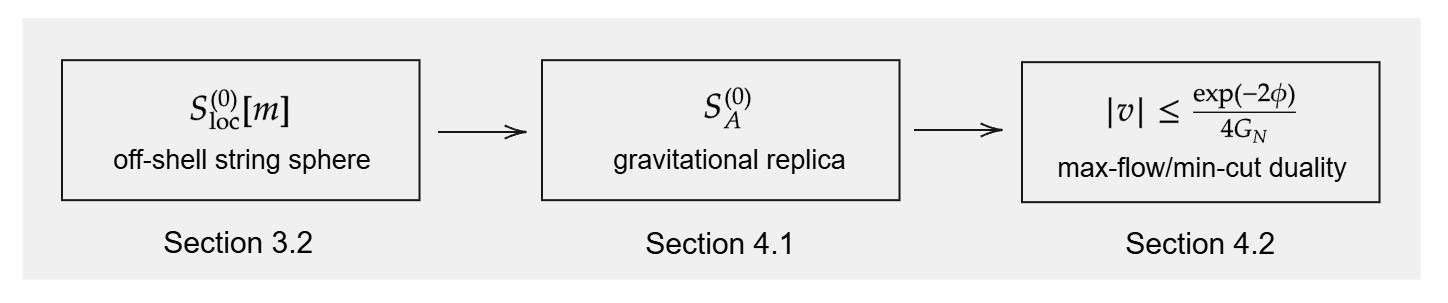}

\noindent We first derive the local entropy capacity from the off-shell
sphere functional. The conical variation of the sphere-derived target
space action assigns a smooth codimension-two surface $m$ the string-frame
functional 
\begin{equation}
S_{\mathrm{loc}}^{\left(0\right)}\left[m\right]=\frac{1}{4G_{N}}\int_{m}e^{-2\phi}dA_{S}=\int_{m}\rho_{S}\left(x\right)dA_{S},\qquad\rho_{S}\left(x\right)=\frac{e^{-2\phi\left(x\right)}}{4G_{N}}.\label{eq:intro-local-capacity}
\end{equation}
The factor $e^{-2\phi}$ originates from the genus-zero closed-string
measure. It also converts the codimension-two string-frame area element
into its Einstein-frame value. The off-shell calculation therefore
identifies $\rho_{S}\left(x\right)$ as the entropy capacity per unit
string frame area.

For a static holographic region $A$, the gravitational replica construction
determines which surface carries this local entropy. Replica regularity
gives the weighted-extremality equation, and semiclassical saddle
dominance selects the weighted minimal cut. Applying weighted flow--cut
duality then yields 
\begin{equation}
S_{A}^{\left(0\right)}=\min_{\substack{\partial m=\partial A\\
m\sim A
}
}\int_{m}\rho_{S}\left(x\right)dA_{S}=\max_{\substack{\nabla_{i}v^{i}=0\\
|v|\leq\rho_{S}
}
}\int_{A}v^{i}n_{i}dA_{S}.\label{eq:intro-weighted-duality}
\end{equation}
This chain gives the string-frame norm bound 
\begin{equation}
|v|\leq\rho_{S}\left(x\right)=\frac{e^{-2\phi\left(x\right)}}{4G_{N}}\label{eq:intro-derived-bound}
\end{equation}
from the sphere-derived entropy density, the gravitational replica
saddle, and convex duality. The off-shell sphere calculation and the
charge flow construction consequently play different roles. The former
fixes the local amount of entropy that a transverse area element can
support. On the other hand, the latter supplies a conserved spatial
realization of that capacity. Their combination applies the same local
density to both a black hole horizon and the weighted RT surface of
a general static boundary region.

\begin{figure}[h]
\centering \includegraphics[scale=0.35]{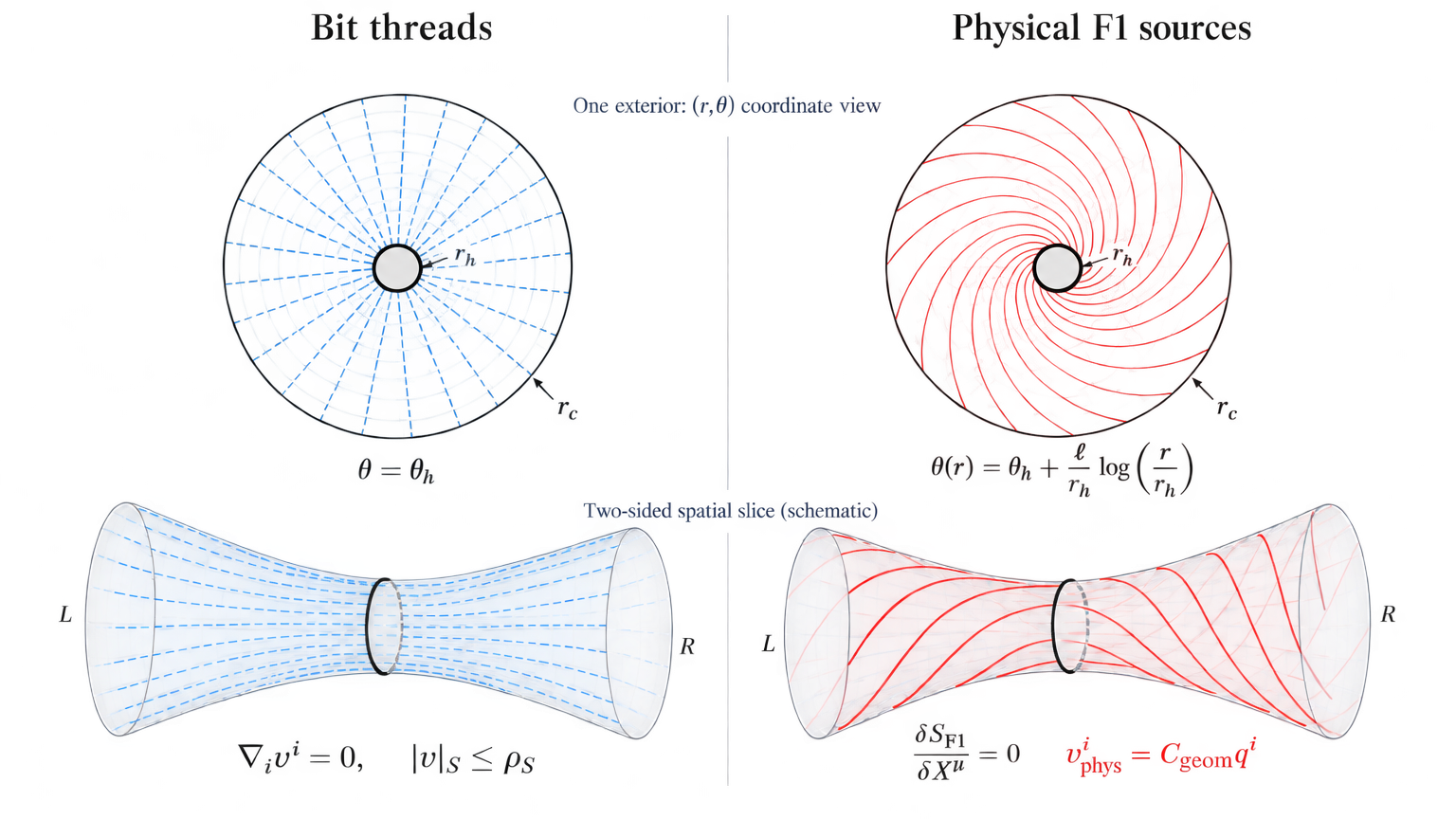} \caption{\label{fig:intro}This figure illustrates the comparison of a radial
bit-thread maximal flow (blue dashed curves) with the flow constructed
from stationary F1 sources (red solid curves) in NS--NS-supported
BTZ. The upper panels show one exterior in $\left(r,\theta\right)$
coordinates. The lower panels show schematic two-sided spatial extensions.
The F1 equations determine the logarithmic spirals, while the off-shell
sphere fixes the entropy capacity used to normalize their charge current.
Both flows attain the same maximal entropy flux, $S_{\mathrm{BH}}^{\left(0\right)}$.}
\end{figure}

We then determine the flow lines from physical F1 dynamics. For nonrotating
BTZ with constant dilaton and NS--NS flux, the radial embedding used
in the original construction is a conserved external source but is
not a free stationary string. The first variation of its Kalb--Ramond
coupling produces an angular force. The regular stationary solution
is instead a logarithmic spiral. Classical string propagation in BTZ
backgrounds was analyzed in \cite{Larsen:1994ah,Ashok:2021ffx}. In
this work, we find the horizon regular branches by solving the probe
F1 equations:
\begin{equation}
\theta\left(r\right)=\theta_{h}+\frac{\ell}{r_{h}}\log\frac{r}{r_{h}}.\label{eq:intro-stationary-profile}
\end{equation}
\emph{This spatial profile is determined directly by the classical
F1 equations in the given background, independently of any prescribed
bit-thread configuration}. The resulting on-shell probe strings therefore
provide a dynamical realization of the entropy-flow lines. For a uniform
ensemble of $\mathcal{N}$ equally oriented strings, its projected
charge density has the constant norm 
\begin{equation}
|q|=\frac{\mathcal{N}}{2\pi r_{h}},\qquad\widehat{q}^{i}=\frac{q^{i}}{|q|},\qquad\nabla_{i}\widehat{q}^{i}=0.\label{eq:intro-unit-current}
\end{equation}
Combining the dynamically selected direction with the independently
derived capacity gives 
\begin{equation}
v_{\mathrm{phys}}^{i}=\rho_{S}\widehat{q}^{i}=C_{\mathrm{geom}}q^{i},\qquad C_{\mathrm{geom}}=\frac{\rho_{S}}{|q|}.\label{eq:intro-physical-flow}
\end{equation}
This flow is conserved, saturates the string-frame norm bound throughout
the BTZ exterior, and becomes normal to the bifurcation circle. Its
flux is 
\begin{equation}
\Phi\left[v_{\mathrm{phys}}\right]=2\pi r_{h}\rho_{S}=\frac{e^{-2\phi_{0}}\left(2\pi r_{h}\right)}{4G_{N}^{\left(3\right)}}=S_{\mathrm{BH}}^{\left(0\right)}.\label{eq:intro-BTZ-result}
\end{equation}
The horizon capacity bounds every admissible flow, so the stationary
F1 current realizes a global maximal flow. Both factors in $C_{\mathrm{geom}}$
are now fixed locally before the entropy flux is integrated: $\rho_{S}$
comes from the off-shell sphere, and $|q|$ comes from the stationary
F1 solution. Fig. (\ref{fig:intro}) illustrates how stationary F1
dynamics provides a physical maximal-flow representative with the
same entropy flux as a radial bit-thread flow.

This result gives the thread picture a concrete physical interpretation.
The string worldsheet is a solution of the source dynamics, its Kalb--Ramond
charge supplies a conserved spacetime current, and the sphere-derived
capacity assigns an entropy weight to that current. The stationary
strings are therefore physical channels carrying the entropic flux
represented by bit threads. For a uniform charge ensemble, 
\begin{equation}
C_{\mathrm{geom}}=\rho_{S}\frac{2\pi r_{h}}{\mathcal{N}}=\frac{S_{\mathrm{BH}}^{\left(0\right)}}{\mathcal{N}},\label{eq:intro-entropy-per-charge}
\end{equation}
so the entropy weight per unit string charge equals the local capacity
multiplied by the portion of the horizon circumference assigned to
that charge. This is the precise sense in which the physical strings
carry entropic information in the present construction.

We finally extend the calculation to BTZ backgrounds with mixed NS--NS
and RR flux, providing a further test of the construction while holding
the metric and dilaton fixed. We use the BTZ quotient of the type
IIB $\mathrm{AdS}_{3}\times S^{3}\times T^{4}$ solution, in which
the NS--NS and RR three-forms support the same metric with complementary
strengths proportional to $\eta$ and $\sqrt{1-\eta^{2}}$ \cite{Cagnazzo:2012se,Hoare:2013pma},
where the parameter $0\leq\eta\leq1$ denotes the NS--NS fraction.
With worldsheet fermions set to zero, the F1 couples directly to the
metric and $B$, while the RR field also contributes to the support
of the background. In this case, the regular stationary branch has
\begin{equation}
\mathcal{D}_{\eta}^{2}=\left(1-\eta^{2}\right)r^{2}+\eta^{2}r_{h}^{2},\qquad\theta_{\eta}^{\prime}=\frac{\eta\ell}{r\mathcal{D}_{\eta}},\qquad|q_{\eta}|=\frac{\mathcal{N}}{2\pi\mathcal{D}_{\eta}}.\label{eq:intro-mixed-current}
\end{equation}
It interpolates between a radial F1 in the pure RR background and
the logarithmic spiral in the pure NS--NS background. For generic
mixed flux, $|q_{\eta}|$ varies with $r$, so its pointwise unit
normalization is not conserved. Instead, the largest constant conversion
coefficient allowed by the local capacity gives 
\begin{equation}
v_{\mathrm{phys},\eta}^{i}=C_{\mathrm{geom}}q_{\eta}^{i},\qquad C_{\mathrm{geom}}=\frac{\rho_{S}}{|q_{\eta}|_{h}}=\frac{2\pi r_{h}\rho_{S}}{\mathcal{N}},\qquad\frac{|v_{\mathrm{phys},\eta}|}{\rho_{S}}=\frac{r_{h}}{\mathcal{D}_{\eta}}\leq1.\label{eq:intro-mixed-flow}
\end{equation}
For every $0\leq\eta\leq1$, the resulting flow $v_{\mathrm{phys},\eta}^{i}$
is conserved, becomes normal to the horizon, and carries the same
maximal entropy flux, $2\pi r_{h}\rho_{S}=S_{\mathrm{BH}}^{\left(0\right)}$.
This example illustrates how physical string dynamics selects different
maximal-flow representatives of the same geometric entropy problem.

In this paper, we show that the off-shell sphere functional determines
the local entropy capacity. The gravitational replica construction
selects the entropy bottleneck. The F1 equations determine physical
charge-current lines. The horizon flux bound proves that the resulting
capacity-normalized current carries the maximal allowed flux. These
results extend the string charge density/bit threads correspondence
by deriving its local entropy capacity and constructing stationary
on-shell F1 representatives of maximal entropy flow. It is worth noting
that all statements in this paper are made at genus zero and two-derivative
order, and the source realizations use classical probe strings and
continuum coarse graining.

The paper is organized as follows. Section \ref{sec:review} reviews
the string charge density/bit-thread correspondence and the normalization
issue in the original BTZ construction. Section \ref{sec:offshell}
derives the two-derivative target-space action and the dilaton-weighted
entropy from the regulated off-shell sphere. Section \ref{sec:norm-bound}
combines this local entropy functional with the gravitational replica
construction and weighted flow--cut duality to obtain the string-frame
norm bound. Section \ref{sec:physical-string-BTZ} constructs the
stationary BTZ strings and proves that their projected current realizes
a maximal entropy flow. Section \ref{sec:extflow} specifies the mixed-flux
bulk and F1 source actions, derives the regular stationary strings
and their projected currents, and proves that the corresponding flows
reproduce the BTZ entropy throughout the NS--NS/RR interpolation.
Section \ref{sec:conclusion} summarizes the results.

\section{String charge density/bit threads correspondence}

\label{sec:review}

In this section, we briefly review the string charge density/bit threads
correspondence introduced in our previous work \cite{Wu:2025qwc}.
The central proposal is that the spatial projection of a conserved
string charge density can provide a physical interpretation of a bit-thread
flow. The construction has two logically distinct ingredients. The
worldsheet coupling to the Kalb--Ramond field determines a conserved
charge current and hence the direction and flux lines of the candidate
flow. Its overall entropy normalization, however, is introduced through
a conversion coefficient $C_{\mathrm{geom}}$. In the original construction
this coefficient is fixed by imposing the bit-thread norm bound and
requiring saturation at the relevant bottleneck. Making this separation
explicit will identify the normalization problem addressed in the
rest of this paper.

\subsection{From the worldsheet current to a conserved spatial flow}

Consider a macroscopic fundamental string propagating in a background
metric, dilaton, and Kalb--Ramond field \cite{Dabholkar:1990yf,Sen:1992yt,Duff:1994an,Lu:2025awh}.
It can be described by the sigma model

\begin{equation}
S_{\sigma}=-\frac{1}{4\pi\alpha^{\prime}}\int d^{2}\sigma\left(\sqrt{-\gamma}\gamma^{mn}\partial_{m}X^{\mathrm{a}}\partial_{n}X^{\mathrm{b}}g_{\mathrm{ab}}+\epsilon^{mn}\partial_{m}X^{\mathrm{a}}\partial_{n}X^{\mathrm{b}}B_{\mathrm{ab}}\right).
\end{equation}
Varying the coupling to $B_{\mathrm{ab}}$ gives the antisymmetric
spacetime current 
\begin{equation}
j^{\mathrm{ab}}=\frac{1}{\sqrt{-g\left(x\right)}}\int d^{2}\sigma\left(\frac{\partial X^{\mathrm{a}}}{\partial\tau}\frac{\partial X^{\mathrm{b}}}{\partial\sigma}-\frac{\partial X^{\mathrm{b}}}{\partial\tau}\frac{\partial X^{\mathrm{a}}}{\partial\sigma}\right)\delta^{\left(3\right)}\left(x-X\left(\tau,\sigma\right)\right).\label{eq:review string current}
\end{equation}
For a closed worldsheet, or locally away from the endpoints of an
open worldsheet, this current obeys 
\begin{equation}
\partial_{\mathrm{a}}\left(\sqrt{-g}j^{\mathrm{ab}}\right)=0.\label{eq:review-spacetime-conservation}
\end{equation}
The current in Eq. (\ref{eq:review string current}) is normalized
so that one consistently oriented fundamental string carries one unit
of Kalb--Ramond charge. 

We now restrict to a static spacetime with vanishing shift and write
\begin{equation}
dS^{2}=-N_{\mathrm{ADM}}^{2}dt^{2}+h_{ij}dx^{i}dx^{j},\qquad\sqrt{-g}=N_{\mathrm{ADM}}\sqrt{h}.
\end{equation}
The projected string charge density on a constant time slice $\Sigma$
is defined by 
\begin{equation}
q^{i}\equiv N_{\mathrm{ADM}}j^{0i}=\frac{\sqrt{-g}}{\sqrt{h}}j^{0i}.\label{eq:review projected current}
\end{equation}
It also obeys
\begin{equation}
\nabla_{i}q^{i}=\frac{1}{\sqrt{h}}\partial_{i}\left(\sqrt{h}q^{i}\right)=0.\label{eq:review-slice-conservation}
\end{equation}
In other words, the projected string charge density automatically
has the conservation property required of a bit-thread flow.

Let us consider the nonrotating BTZ black hole supported by NS--NS
flux \cite{Banados:1992wn,Horowitz:1993jc}. In string-frame coordinates
adapted to one exterior region, 
\begin{align}
dS_{\mathrm{BTZ}}^{2} & =-\frac{r_{h}^{2}}{\ell^{2}}\sinh^{2}\rho dt^{2}+\ell^{2}d\rho^{2}+r_{h}^{2}\cosh^{2}\rho d\theta^{2}, & \theta & \sim\theta+2\pi,\\
B_{t\theta} & =-\frac{r_{h}^{2}}{\ell}\sinh^{2}\rho, & \phi & =\phi_{0}.\label{eq:review-BTZ-background}
\end{align}
The bifurcation circle is at $\rho=0$ and has string-frame circumference
$2\pi r_{h}$.

\begin{figure}[h]
\begin{centering}
\includegraphics[scale=0.4]{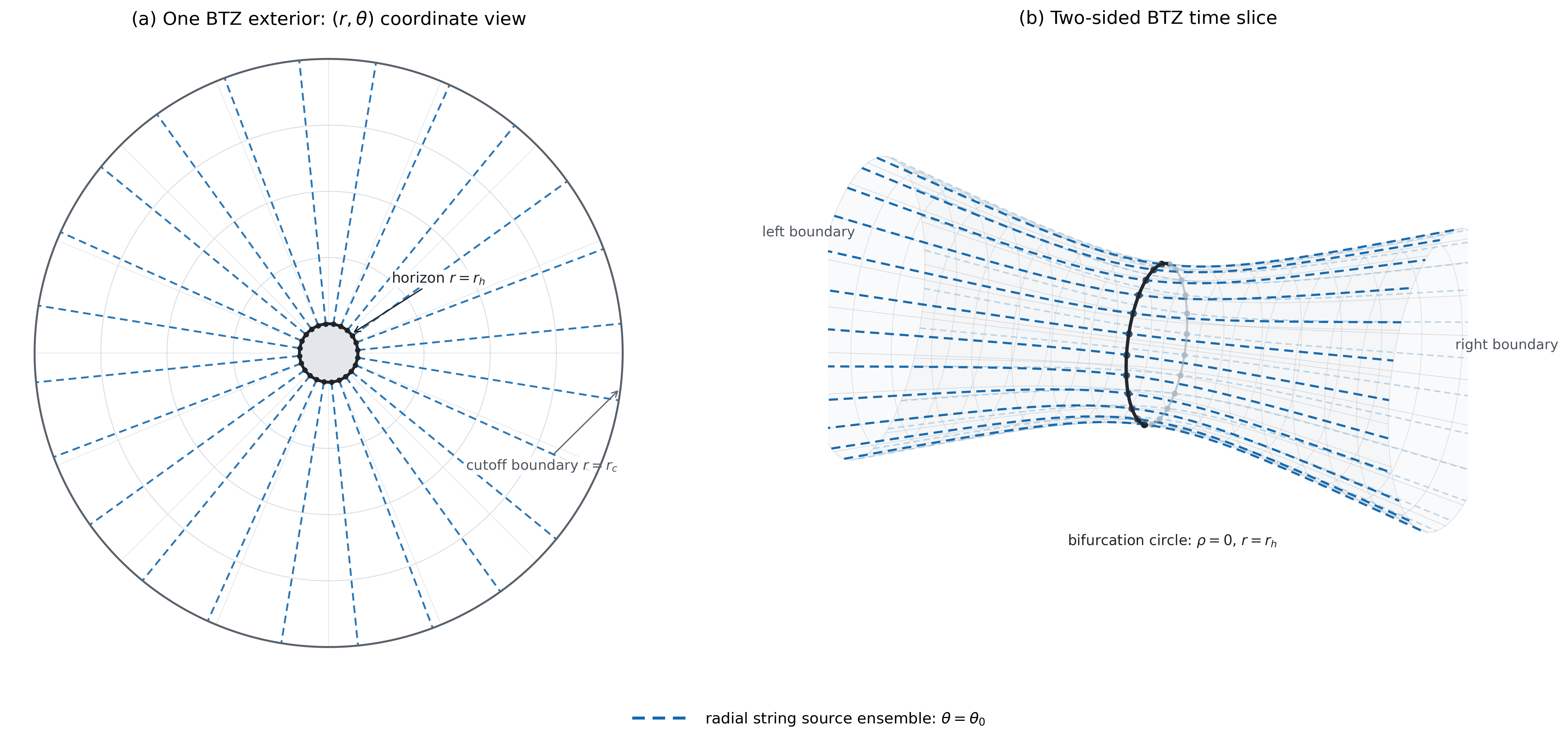}
\par\end{centering}
\caption{\label{fig:old source}This figure illustrates the radial string-source
ensembles on static BTZ slices. Blue dashed lines represent prescribed
radial string sources with $X^{\theta}=\theta_{0}$, uniformly distributed
in the angular coordinate. (a) A single BTZ exterior shown in $\left(r,\theta\right)$
coordinates, between the horizon $r=r_{h}$ and the radial cutoff
$r=r_{c}$. (b) The corresponding ensemble extended across the two-sided
BTZ slice, with the bifurcation circle at the throat.}
\end{figure}

In the original construction, a radial source embedding is prescribed
in static gauge:
\begin{equation}
X^{t}=\tau,\qquad X^{\rho}=\sigma,\qquad X^{\theta}=\theta_{0}.\label{eq:review-radial-embedding}
\end{equation}
For one string, Eq. (\ref{eq:review string current}) gives 
\begin{equation}
j_{\mathrm{BTZ}}^{0\rho}=\frac{1}{\sqrt{-g}}\delta_{2\pi}\left(\theta-\theta_{0}\right).
\end{equation}
For $\mathcal{N}$ equally oriented strings distributed uniformly
around the angular circle, coarse-graining replaces the periodic delta
function by its angular density $\mathcal{N}/\left(2\pi\right)$.
Specifically, the corresponding current for $\mathcal{N}$ equally
oriented strings is given by

\begin{equation}
j_{\mathrm{BTZ}}^{0\rho}=\frac{1}{\sqrt{-g}}\stackrel[k=0]{\mathcal{N}-1}{\sum}\delta_{2\pi}\left(\theta-\frac{2\pi k}{\mathcal{N}}\right).
\end{equation}
The illustration of this configuration can be found in Fig. (\ref{fig:old source}).
 Denoting the angular separation between any two adjacent strings
as $\theta_{s}=2\pi/\mathcal{N}$, the string charge density can be
rewritten as

\begin{equation}
j_{\mathrm{BTZ}}^{0\rho}=\frac{1}{\sqrt{-g}}\frac{1}{\theta_{s}}\left[\theta_{s}\stackrel[k=0]{\mathcal{N}-1}{\sum}\delta_{2\pi}\left(\theta-k\theta_{s}\right)\right]=\frac{1}{\sqrt{-g}}\frac{\mathcal{N}}{2\pi}\left[\theta_{s}\stackrel[k=0]{\mathcal{N}-1}{\sum}\delta_{2\pi}\left(\theta-k\theta_{s}\right)\right].
\end{equation}
Hence, coarse-graining over angular scales larger than the separation
between neighboring strings gives
\begin{equation}
j_{\mathrm{BTZ}}^{0\rho}=\frac{1}{\sqrt{-g}}\frac{\mathcal{N}}{2\pi}=\frac{2}{r_{h}^{2}\sinh\left(2\rho\right)}\frac{\mathcal{N}}{2\pi},\qquad j^{0\theta}=0.\label{eq:review-BTZ-j}
\end{equation}
The lapse and the induced spatial metric are 
\begin{equation}
N_{\mathrm{ADM}}=\frac{r_{h}}{\ell}\sinh\rho,\qquad h_{ij}dx^{i}dx^{j}=\ell^{2}d\rho^{2}+r_{h}^{2}\cosh^{2}\rho d\theta^{2}.
\end{equation}
The projected string charge density is therefore 
\begin{equation}
q_{\mathrm{BTZ}}^{i}=N_{\mathrm{ADM}}j^{0i}=\frac{\mathcal{N}}{2\pi\ell r_{h}\cosh\rho}\left(1,0\right),\qquad i=\rho,\theta,\label{eq:review-BTZ-q}
\end{equation}
and its norm is 
\begin{equation}
|q_{\mathrm{BTZ}}|=\sqrt{h_{ij}q_{\mathrm{BTZ}}^{i}q_{\mathrm{BTZ}}^{j}}=\frac{\mathcal{N}}{2\pi r_{h}\cosh\rho}.\label{eq:review-BTZ-q-norm}
\end{equation}
The candidate bit-thread/entropy flow can be given by using the dimensionless
conversion coefficient $C_{\mathrm{geom}}^{\mathrm{BTZ}}$:
\begin{equation}
v_{\mathrm{BTZ}}^{i}=C_{\mathrm{geom}}^{\mathrm{BTZ}}q_{\mathrm{BTZ}}^{i},
\end{equation}
which is divergenceless, and its norm is 
\begin{equation}
|v_{\mathrm{BTZ}}|=C_{\mathrm{geom}}^{\mathrm{BTZ}}\frac{\mathcal{N}}{2\pi r_{h}\cosh\rho}.\label{eq:review-BTZ-v-norm}
\end{equation}
Because $\left(1/\cosh\rho\right)\leq1$, the norm is maximal at the
bifurcation circle. Imposing the string-frame norm bound and requiring
the flow to saturate it at $\rho=0$ gives \cite{Freedman:2016zud,Headrick:2017ucz,Wu:2025qwc}
\begin{equation}
C_{\mathrm{geom}}^{\mathrm{BTZ}}\frac{\mathcal{N}}{2\pi r_{h}}=\frac{e^{-2\phi_{0}}}{4G_{N}^{\left(3\right)}},
\end{equation}
and therefore 
\begin{equation}
C_{\mathrm{geom}}^{\mathrm{BTZ}}=\frac{2\pi r_{h}}{\mathcal{N}}\frac{e^{-2\phi_{0}}}{4G_{N}^{\left(3\right)}}=\frac{e^{-2\phi_{0}}\pi r_{h}}{2G_{N}^{\left(3\right)}\mathcal{N}}.\label{eq:review-BTZ-C}
\end{equation}
In the convention $\phi_{0}=0$, this reduces to the coefficient used
in the original BTZ calculation.

The resulting bit-thread/entropy flow is 
\begin{equation}
v_{\mathrm{BTZ}}^{i}=\frac{e^{-2\phi_{0}}}{4G_{N}^{\left(3\right)}\ell\cosh\rho}\left(1,0\right).\label{eq:review-BTZ-flow}
\end{equation}
For a circle of constant $\rho$, the outward unit normal covector
and the induced line element are 
\begin{equation}
n_{i}dx^{i}=\ell d\rho,\qquad dA_{S}=r_{h}\cosh\rho d\theta.
\end{equation}
Its flux is consequently independent of $\rho$: 

\begin{equation}
\Phi\left[v_{\mathrm{BTZ}}\right]=\int_{0}^{2\pi}v_{\mathrm{BTZ}}^{i}n_{i}dA_{S}=\frac{e^{-2\phi_{0}}\left(2\pi r_{h}\right)}{4G_{N}^{\left(3\right)}}=S_{\mathrm{BH}}^{\left(0\right)}.
\end{equation}
Thus the projected string charge density determines a conserved set
of radial flow lines, while the norm-bound saturation determines how
much entropy flux is assigned to those lines. The two requirements
combine to reproduce the BTZ entropy.

The correspondence was subsequently extended to ten-dimensional three
charge black holes and tested under string dualities. In the Type
IIB S-duality analysis of the F1--NS5--P and D1--D5--P systems
\cite{Wu:2026qha}, an F1 current is mapped to a D1 current. The local
metric, dilaton, charge current, and flow norm change between the
two descriptions, but the maximal integrated entropy flux is preserved.
In the T-duality analysis \cite{Wu:2026kci}, a radial F1 source remains
an F1 source along the NS--NS duality chain, whereas a radial D1
source transverse to the duality circle becomes a D2 source wrapped
on the dual circle. Contracting the projected D2 bivector with the
normalized one-form along that circle produces an effective conserved
vector flow. Again, the local representatives change, while the dilaton-weighted
transverse capacity and the integrated maximal entropy flux remain
invariant. These results provide nontrivial duality checks that string
and brane charge currents can furnish different microscopic representatives
of the same entropy flow.

However, two ingredients of the original construction require a firmer
foundation from string theory.
\begin{enumerate}
\item \textbf{Independent determination of the entropy capacity.} In the
original identification, 
\begin{equation}
v_{\mathrm{BTZ}}^{i}=C_{\mathrm{geom}}^{\mathrm{BTZ}}q_{\mathrm{BTZ}}^{i},
\end{equation}
the coefficient $C_{\mathrm{geom}}^{\mathrm{BTZ}}$ is fixed by imposing
the bit-thread norm bound and requiring saturation at the horizon.
The entropy normalization therefore relies on the geometric capacity
constraint whose string-theoretic origin we seek to explain. Our aim
is to derive the entropy capacity from the off-shell string sphere
functional and establish the corresponding string-frame norm bound
through the replica construction and flow--cut duality. Combining
this independently determined capacity with the dynamically selected
string current then fixes the conversion coefficient before evaluating
the total entropy flux, allowing the BTZ entropy to follow from the
construction without matching the final flux to a prescribed black-hole
entropy.
\item \textbf{Dynamical selection of the string embeddings.} In the original
BTZ construction, the parallel radial strings are introduced through
embeddings described above. Although their projected charge density
is conserved and reproduces the desired flow, current conservation
alone does not establish that these embeddings satisfy the stationary
fundamental-string equations of motion. This motivates us to determine
the string arrangement directly from the F1 dynamics and then establish
whether its projected current realizes a maximal entropy flow. In
this way, the string configuration is selected by the worldsheet equations
rather than prescribed using a known bit-thread solution.
\end{enumerate}

\section{Black hole entropy from off-shell strings}

\label{sec:offshell}

The purpose of this section is to determine the local entropy capacity
independently of the F1 source construction. We first review Tseytlin's
regulated sphere prescription \cite{Tseytlin:1987ww,Tseytlin:1988tv,Ahmadain:2022tew,Ahmadain:2022eso}
and show how the logarithmic cutoff dependence of the genus-zero worldsheet
functional produces the two-derivative string-frame effective action.
We then evaluate the same off-shell functional on a regulated conical
target-space geometry. The contribution linear in the conical deficit
localizes on the bifurcation surface $\Gamma_{H}$ and gives 
\begin{equation}
S_{\mathrm{BH}}^{(0)}=\frac{1}{4G_{N}}\int_{\Gamma_{H}}e^{-2\phi}dA_{S}.\label{eq:section3-roadmap-entropy}
\end{equation}
This result identifies 
\begin{equation}
\rho_{S}\left(x\right)=\frac{e^{-2\phi(x)}}{4G_{N}}\label{eq:section3-roadmap-capacity}
\end{equation}
as the entropy capacity per unit string-frame area. Note that the
fixed-cone calculation determines this density for a given surface.
It does not yet identify the surface that computes the entropy of
a general boundary region. Section \ref{sec:norm-bound} provides
that selection through the gravitational replica construction and
then converts the weighted minimal cut into a maximal flow.

\subsection{Brief review of low energy effective action from off-shell strings}

Let 
\begin{equation}
\Psi=\left\{ g_{\mu\nu},B_{\mu\nu},\phi,\ldots\right\} 
\end{equation}
denote the target-space fields. We use $g_{\mu\nu}$ for the Euclidean
target space string-frame metric and $\gamma_{ab}$ for the worldsheet
metric. On a closed Euclidean worldsheet $\mathcal{W}$, the universal
NS--NS couplings are \cite{Fradkin:1985ys,Callan:1985ia}
\begin{align}
I_{\sigma}\left[X;\Psi\right]=\frac{1}{4\pi\alpha^{\prime}}\int_{\mathcal{W}}d^{2}\sigma\bigg[ & \sqrt{\gamma}\gamma^{ab}g_{\mu\nu}\left(X\right)\partial_{a}X^{\mu}\partial_{b}X^{\nu}+i\epsilon^{ab}B_{\mu\nu}\left(X\right)\partial_{a}X^{\mu}\partial_{b}X^{\nu}\nonumber \\
 & +\alpha^{\prime}\sqrt{\gamma}\phi\left(X\right)R^{\left(2\right)}[\gamma]\bigg],\label{eq:sigma-model-replica-section}
\end{align}
where $\epsilon^{ab}$ is the antisymmetric density, with $\epsilon^{12}=+1$.
The Fradkin--Tseytlin term already contains the dilaton dependence.
Writing $\phi=\phi_{\infty}+\varphi$ and $g_{s}=e^{\phi_{\infty}}$,
its constant part contributes 
\begin{equation}
\exp\left[-\frac{\phi_{\infty}}{4\pi}\int_{\mathcal{W}}\sqrt{\gamma}R^{\left(2\right)}\right]=e^{-\phi_{\infty}\chi\left(\mathcal{W}\right)}.\label{eq:dilaton-euler-factor}
\end{equation}
For the sphere, $\chi\left(S^{2}\right)=2$, and this factor is $g_{s}^{-2}$.
When the nonconstant dilaton is evaluated on the worldsheet zero mode
$Y$, the complete sphere weight is 
\begin{equation}
g_{s}^{-2}e^{-2\varphi\left(Y\right)}=e^{-2\phi\left(Y\right)}.\label{eq:full-dilaton-sphere-weight}
\end{equation}
On the sphere, $\chi=2$, so the dilaton coupling gives the factor
$e^{-2\phi}$. For a constant dilaton, this gives the tree-level scaling
$I_{0}=O\left(g_{s}^{-2}\right)$. The same dilaton weight appears
in the string-frame effective action and in the local entropy density
in Eq. (\ref{eq:section3-roadmap-capacity}). We retain $B_{\mu\nu}$
in the field content for generality, although a smooth $B$ field
produces no contribution localized at the conical tip at this order;
the leading entropy term comes from the curvature coupling weighted
by the dilaton.

To make the prescription explicit, we expand locally in theory space
around a reference worldsheet CFT \cite{Ahmadain:2022tew,Ahmadain:2022eso,Ahmadain:2024hdp}.
Let ${\cal O}_{a}$ be scalar operators of weights $\left(h_{a},h_{a}\right)$
in the gauge-fixed worldsheet theory, with the ghost sector included
in the functional measure. The perturbed worldsheet action is
\begin{equation}
I_{{\rm QFT}}=I_{{\rm CFT}}+\sum_{a}\epsilon^{2\left(h_{a}-1\right)}\omega^{a}V_{a},\qquad V_{a}=\int_{S^{2}}\mathrm{d}^{2}z\sqrt{g}{\cal O}_{a}\left(z,\bar{z}\right),\label{eq:regulated-worldsheet-deformation}
\end{equation}
where $\Delta_{a}=2h_{a}$ is the total scaling dimension, and the
cutoff factor makes $\omega^{a}$ dimensionless. A physical integrated
closed-string vertex is a $\left(1,1\right)$ operator satisfying
the usual physical-state conditions. Operators with $h_{a}\neq1$
describe off-shell external deformations. The marginal operators can
also develop nonzero beta functions through interactions. Based on
this worldsheet action, for the massless NS--NS fields, the couplings
$\omega^{a}$ denote the mode coefficients of massless fields $g_{\mu\nu}$,
$B_{\mu\nu}$, and $\phi$. Infinitesimally, the regulated sphere
functional including matter and ghost can be expanded by the number
$n$ of integrated insertions:
\begin{align}
K_{g}\left(\tau\right) & =\sum_{n=0}^{\infty}K_{g,n}\left(\tau\right),\label{eq:K0-functional-expansion}\\
K_{g,n} & =\sum_{a_{1}\ldots a_{n}}\left(\prod^{n}\epsilon^{2\left(h_{a}-1\right)}\omega^{a}\right)\frac{1}{n!}\left\langle \!\left\langle V_{a_{1}}\cdots V_{a_{n}}\right\rangle \!\right\rangle _{{\rm CFT}}.\label{eq:K0N-regulated-correlator}
\end{align}

\noindent where $\tau$ denotes the worldsheet moduli and it becomes
absent when genus $g=0$ sphere. This sphere can be written as $K_{0}=\sum_{n}K_{0,n}$.
In this expansion, all $n$ vertex operators $V_{a}$ are integrated,
including the case $g=0$. Moreover, the operator collisions are removed
by introducing the hard disks of radius $\epsilon$ around each of
vertex operator insertions. So the integration region obeys
\begin{equation}
d\left(z_{i},z_{j}\right)\geq2\epsilon\qquad\left(i\neq j\right).\label{eq:hard-disk-regulator-section}
\end{equation}

Now, let us recall Tseytlin's two sphere prescriptions \cite{Tseytlin:1987ww,Tseytlin:1988tv,Ahmadain:2022eso},
which act separately on every $K_{0,n}$:
\begin{align}
\mathbf{T1}:\qquad A_{0,n} & =\frac{\partial K_{0,n}}{\partial\log\epsilon},\label{eq:T1-amplitude}\\
\mathbf{T2}:\qquad A_{0,n} & =\left(\frac{\partial}{\partial\log\epsilon}+\frac{1}{2}\frac{\partial^{2}}{\left(\partial\log\epsilon\right)^{2}}\right)K_{0,n}.\label{eq:T2-amplitude}
\end{align}
Resumming the connected sphere amplitudes in $n$ defines
\begin{equation}
-I_{0}=Z_{0}=\sum_{n=0}^{\infty}A_{0,n}.\label{eq:Tseytlin-prescription-section}
\end{equation}
Thus, the $n$-independent actions is given by \cite{Ahmadain:2022eso}
\begin{eqnarray}
I_{0}^{\left(\mathbf{T1}\right)} & = & -\left(\frac{\partial}{\partial\log\epsilon}\right)K_{0},\nonumber \\
I_{0}^{\left(\mathbf{T2}\right)} & = & -\left(\frac{\partial}{\partial\log\epsilon}+\frac{1}{2}\frac{\partial^{2}}{\left(\partial\log\epsilon\right)^{2}}\right)K_{0}.
\end{eqnarray}
Moreover, this action also can be written in terms of $\beta$ functions,
since the differentiating $\omega^{a}$ with respect to $\log\epsilon$
corresponds to $\beta$ funtions in the RG flow:
\begin{equation}
\beta^{a}:=-\frac{d\omega^{a}}{d\log\epsilon}.\label{eq:AW-beta-convention}
\end{equation}
In this case, the \textbf{T1} action can be rewritten as
\begin{equation}
I_{0}^{\left(\mathbf{T1}\right)}=-\frac{\partial K_{0}}{\partial\log\epsilon}=\sum_{a}\beta^{a}\frac{\partial K_{0}}{\partial\omega^{a}}.\label{eq:T1-action-beta-form}
\end{equation}
And the \textbf{T2} action can be given by
\begin{equation}
I_{0}^{\left(\mathbf{T2}\right)}=I_{0}^{\left(\mathbf{T1}\right)}-\frac{1}{2}\left(\sum_{a,b}\beta^{a}\beta^{b}\frac{\partial^{2}K_{0}}{\partial\omega^{a}\partial\omega^{b}}+\sum_{a,b}\beta^{b}\frac{\partial\beta^{a}}{\partial\omega^{b}}\frac{\partial K_{0}}{\partial\omega^{a}}\right).
\end{equation}
It is worth noting that every term of action is proportional to at
least one beta function. To obtain the equations of motion and low
energy effective action, it is sufficient to use the \textbf{T1} prescription.
Indeed, after the cancellation of the power-law divergences between
the matter measure and the Gaussian determinant, and the cancellation
of the remaining critical-dimension contribution against the ghosts,
the cutoff dependence of $K_{0}$ is linear in $\log\epsilon$ at
$O\left(\alpha^{\prime}\right)$. It implies
\begin{equation}
I_{0}^{\left(\mathbf{T2}\right)}=I_{0}^{\left(\mathbf{T1}\right)}+O\left(\alpha^{\prime2}\right).
\end{equation}
The \textbf{T2} prescription becomes essential when the bosonic closed-string
tachyon is included, since its additional second-derivative term automatically
removes the spurious contribution proportional to the worldsheet cosmological-constant
divergence $\epsilon^{-2}$. In our case, we do not consider the tachyon,
which implies that the simpler \textbf{T1} prescription is sufficient
at this order.

Now, let us see how the equations of motion and the corresponding
low-energy effective action are obtained using Ahmadain and Wall's
method \cite{Ahmadain:2022tew,Ahmadain:2022eso,Ahmadain:2024hdp}.
We first consider the bosonic nonlinear sigma model on a compact Euclidean
worldsheet $\mathcal{W}$, 
\begin{equation}
Z^{\left(B\right)}\left(g,\phi;\epsilon\right)=\int\left[dX\right]e^{-I_{{\rm QFT}}\left[X\right]},
\end{equation}
where $\left[dX\right]=\underset{z}{\prod}dx\left(z\right)\sqrt{g\left(X\left(z\right)\right)}$
and
\begin{equation}
I_{{\rm QFT}}\left[X\right]=\frac{1}{4\pi\alpha^{\prime}}\int_{\mathcal{W}}d^{2}z\sqrt{\gamma}\left[g_{\mu\nu}\left(X\right)\partial_{a}X^{\mu}\partial^{a}X^{\nu}+\alpha'R^{\left(2\right)}\left[\gamma\right]\phi\left(X\right)\right].
\end{equation}
Here $g_{\mu\nu}\left(X\right)$ is the spacetime metric and $\phi\left(X\right)$
is the dilaton. For a spherical worldsheet, the regulated functional
entering Tseytlin's \textbf{T1} prescription is 
\begin{equation}
K_{0}\left(\epsilon\right)=Z^{\left(B\right)}Z_{{\rm ghost}}.
\end{equation}
If we can show the two-derivative part of the regulated functional
has the schematic form
\begin{equation}
K_{0}\left(\epsilon\right)=K_{0}^{\left(0\right)}+\left(\log\epsilon\right)\int\mathrm{d}^{D}Y{\cal L}_{\log}\left(Y\right)+\text{finite}+O\left(\alpha^{\prime2}\right),\label{eq:log-coefficient-goal}
\end{equation}

\noindent then the \textbf{T1} action gives
\begin{equation}
I_{0}^{\left(\mathbf{T1}\right)}=-\left(\frac{\partial}{\partial\log\epsilon}\right)K_{0}=-\int\mathrm{d}^{D}Y{\cal L}_{\log}\left(Y\right).
\end{equation}
It gives us low energy effective action. Therefore, the main aim of
following calculations is to obtain $K_{0}\left(\epsilon\right)$
and identify ${\cal L}_{\log}\left(Y\right)$ explicitly.

To approach this aim, we split the embedding field $X^{\mu}\left(z\right)$
into its constant and nonconstant parts,
\begin{equation}
X^{\mu}\left(z\right)=Y^{\mu}+\eta^{\mu}\left(z\right),\qquad\int_{\mathcal{W}}\mathrm{d}^{2}z\sqrt{\gamma}\eta^{\mu}\left(z\right)=0,\label{eq:worldsheet-zero-mode-split}
\end{equation}
where $Y^{\mu}$ contains zero modes and $\eta^{\mu}$ contains nonzero
modes. To understand this mode decomposition, let us recall the compact,
connected worldsheet $\mathcal{W}$, the target space can be expanded
as
\begin{equation}
X^{\mu}\left(z\right)=\sum_{m=0}^{\infty}x_{m}^{\mu}\varphi_{m}\left(z\right),
\end{equation}
where
\begin{equation}
\Delta\varphi_{m}=\lambda_{m}\varphi_{m},\qquad\int_{\mathcal{W}}\mathrm{d}^{2}z\sqrt{\gamma}\varphi_{m}\varphi_{n}=\delta_{mn}.\label{eq:AW-mode-basis}
\end{equation}
For the non-bounded compact worldsheet, we have
\begin{equation}
\lambda_{m}=\int_{\mathcal{W}}\mathrm{d}^{2}z\sqrt{\gamma}\nabla_{A}\varphi_{m}\nabla^{A}\varphi_{n}\geq0.
\end{equation}
Moreover, if $\lambda_{0}=0$, it implies $\nabla_{A}\varphi_{0}=0$
and $\varphi_{0}$ must be a constant. The normalization condition
gives
\begin{equation}
1=\int_{\mathcal{W}}\mathrm{d}^{2}z\sqrt{\gamma}\varphi_{0}^{2}=V\varphi_{0}^{2},\qquad V\equiv\int_{\mathcal{W}}\mathrm{d}^{2}z\sqrt{\gamma},
\end{equation}
which gives $\varphi_{0}=1/\sqrt{V}$. Therefore, it is easy to get
\begin{eqnarray}
X^{\mu}\left(z\right) & = & x_{0}^{\mu}\varphi_{0}\left(z\right)+\sum_{m\neq0}x_{m}^{\mu}\varphi_{m}\left(z\right)\nonumber \\
 & = & \frac{x_{0}^{\mu}}{\sqrt{V}}+\sum_{m\neq0}x_{m}^{\mu}\varphi_{m}\left(z\right)\nonumber \\
 & \equiv & Y^{\mu}+\eta^{\mu}\left(z\right).
\end{eqnarray}
Now, the regulated scalar propagator for non-zero modes can be given
by
\begin{equation}
D\left(z,z^{\prime};\epsilon\right)=\sum_{m\neq0}\frac{e^{-\epsilon^{2}\lambda_{m}}}{\lambda_{m}}\varphi_{m}\left(z\right)\varphi_{m}\left(z^{\prime}\right),
\end{equation}
Acting with $\partial/\partial\epsilon^{2}$ gives 
\begin{equation}
\frac{\partial D\left(z,z^{\prime};\epsilon\right)}{\partial\epsilon^{2}}=-\sum_{m\neq0}e^{-\epsilon^{2}\lambda_{m}}\varphi_{m}\left(z\right)\varphi_{m}\left(z^{\prime}\right),
\end{equation}
where the heat kernel is given by \cite{Vassilevich:2003xt}
\begin{equation}
K\left(z,z^{\prime};\epsilon\right)=\sum_{m\geq0}e^{-\epsilon^{2}\lambda_{m}}\varphi_{m}\left(z\right)\varphi_{m}\left(z^{\prime}\right).
\end{equation}
Since $\varphi_{0}\left(z\right)\varphi_{0}\left(z^{\prime}\right)=1/V$,
the derivative identity, valid at finite regulator, is
\begin{equation}
\frac{\partial D\left(z,z;\epsilon\right)}{\partial\epsilon^{2}}=-K\left(z,z;\epsilon\right)+\frac{1}{V},
\end{equation}
where the coincident heat kernel has the small-$\epsilon$ expansion
\begin{equation}
K\left(z,z;\epsilon\right)=\frac{1}{4\pi\epsilon^{2}}+\frac{1}{24\pi}R^{\left(2\right)}+O\left(\epsilon^{2}\right).
\end{equation}
Integrating with respect to $\epsilon^{2}$ gives the coincident propagator
\begin{equation}
D\left(z,z;\epsilon\right)=-\frac{1}{4\pi}\log\epsilon^{2}+h=-\frac{1}{2\pi}\log\epsilon+h.
\end{equation}
Therefore, the term $\log\epsilon$ that we want comes from the UV
singularity of coincident propagator. After the regularization, the
number of all modes $N$ can be obtained through
\begin{equation}
N\equiv\mathrm{Tr}e^{-\epsilon^{2}\Delta}=\int\mathrm{d}^{2}z\sqrt{\gamma}K\left(z,z;\epsilon\right)=\frac{V}{4\pi\epsilon^{2}}+\frac{1}{6}\chi+O\left(\epsilon^{2}\right),\label{eq:AW-number-of-modes}
\end{equation}
Equivalently, the regulated number of propagating nonzero modes is
\begin{equation}
N^{\prime}=\mathrm{Tr}^{\prime}e^{-\epsilon^{2}\Delta}=N-1.
\end{equation}

Using the full dilaton $\phi=\phi_{\infty}+\varphi$ throughout, the
partition function takes the form
\begin{eqnarray}
Z^{\left(B\right)} & = & Z_{f}^{\left(0\right)}\int d^{D}Y\sqrt{g}e^{-\chi\phi}Z_{D}^{\left(1\right)}Z_{g}^{\left(1\right)}Z_{\mathcal{M}}^{\left(1\right)}\left(Y\right)\nonumber \\
 & = & Z_{f}^{\left(0\right)}\int d^{D}Y\sqrt{g}e^{-\chi\phi}\left(1+\delta Z_{\phi}+\delta Z_{g}+\delta Z_{\mathcal{M}}+O\left(\alpha^{\prime2}\right)\right),
\end{eqnarray}
where $Z_{D}^{\left(1\right)}$ is the 1-loop dilaton multiplicative
corrections, $Z_{g}^{\left(1\right)}$ is the 2-loop graviton multiplicative
corrections, and $Z_{\mathcal{M}}^{\left(1\right)}$ is the measure
factor. We now split the embedding field into its constant mode and
nonzero-mode fluctuations $X^{\mu}\left(z\right)=Y^{\mu}+\eta^{\mu}\left(z\right)$
and Taylor-expand the background fields about $Y^{\mu}$. Expanding
both the interaction exponential and the regulated functional measure,
and then evaluating the resulting Gaussian expectation values with
respect to the quadratic action for $\eta^{\mu}$, gives, through
$O\left(\alpha^{\prime}\right)$ and to second order in target-space
derivatives,
\begin{equation}
\delta Z_{\phi}=-\pi\chi\alpha^{\prime}D\left(z,z;\epsilon\right)g^{\mu\nu}\partial_{\mu}\partial_{\nu}\phi.
\end{equation}
On the other hand, the metric and measure sectors $Z_{g}^{\left(1\right)}$
and $Z_{\mathcal{M}}^{\left(1\right)}$ generate several intermediate
contractions involving $\partial_{\lambda}\partial_{\rho}g_{\mu\nu}$
and $\partial_{\rho}g_{\mu\nu}\partial_{\lambda}g_{\alpha\beta}$.
The explicit forms and coefficients are given in ref. \cite{Ahmadain:2022eso}.
Using
\begin{equation}
D\left(z,z;\epsilon\right)=-\frac{1}{2\pi}\log\epsilon+h,\qquad N-N^{\prime}=1,
\end{equation}
and combining the dilaton, metric, and measure contributions, one
obtains
\begin{equation}
\delta Z_{\phi}+\delta Z_{g}+\delta Z_{\mathcal{M}}\doteq\frac{\alpha^{\prime}}{2}\left(\log\epsilon+h\right)\left(R+\chi^{2}\partial_{\mu}\phi\partial^{\mu}\phi\right).
\end{equation}
Here $\doteq$ denotes equality under the weighted target-space integral
with measure $d^{D}Y\sqrt{g}e^{-\chi\phi}$. Finally, we have
\begin{equation}
Z^{\left(B\right)}=g_{s}^{-\chi}Z_{f}^{\left(0\right)}\int d^{D}Y\sqrt{g}e^{-\chi\phi}\left[1+\frac{\alpha^{\prime}}{2}\left(\log\epsilon+h\right)\left(R+\chi^{2}\partial_{\mu}\phi\partial^{\mu}\phi\right)+O\left(\alpha^{\prime2}\right)\right].
\end{equation}
Multiplying by the ghost contribution $K_{0}\left(\epsilon\right)=Z^{\left(B\right)}Z_{{\rm ghost}}$
and setting $D=26$ cancels the field-independent central-charge contribution
of the free bosons. For $\chi=2$, the remaining two-derivative logarithm
$\log\epsilon$ of $Z_{f}^{\left(0\right)}$ gives the critical bosonic
string action

\begin{equation}
I_{0}=-\frac{\partial K_{0}}{\partial\log\epsilon}\ \propto\ -\alpha^{\prime}\int\mathrm{d}^{26}Y\sqrt{g}e^{-2\phi}\left(R+2\nabla^{2}\phi\right).\label{eq:AW-explicit-bosonic-action}
\end{equation}
Using covariant total derivatives,
\begin{equation}
\int\sqrt{g}e^{-2\phi}\nabla^{2}\phi=2\int\sqrt{g}e^{-2\phi}\left(\nabla\phi\right)^{2}+\int_{\partial\mathcal{M}}\sqrt{h}e^{-2\phi}n^{\mu}\nabla_{\mu}\phi,\label{eq:dilaton-integration-by-parts}
\end{equation}
the bulk integrand becomes $R+4\left(\nabla\phi\right)^{2}$ as expected.

Finally, we need to fix the normalization constant. The proportionality
sign in (\ref{eq:AW-explicit-bosonic-action}) implies that the heat
kernel calculation only determines the target space structure and
its dependence on $g_{s}$ and $\alpha^{\prime}$, but does not provide
all numerical measure and kinematic factors. The overall normalization
is fixed independently by expanding $g_{\mu\nu}$ about flat space
and matching the resulting graviton kinetic term and interaction vertices
to the standard genus-zero graviton scattering amplitude. Since the
Tseytlin sphere prescription reproduces the usual tree-level closed
string S-matrix, this matching fixes the coefficient $e^{-2\phi_{\infty}}/\left(16\pi G_{N}\right)$
of the string-frame graviton kinetic term, with the dilaton dependence
retained explicitly in the convention used here. Restoring this normalization,
one may write
\begin{equation}
I_{0}=-\frac{1}{16\pi G_{N}}\int d^{26}Y\sqrt{g}e^{-2\phi}\left(R+2\nabla^{2}\phi+O\left(\alpha^{\prime}\right)\right).\label{eq:AW action}
\end{equation}
Thus the overall gravitational coefficient is fixed by the normalized
sphere graviton amplitude, independently of entropy matching. In our
full-dilaton convention, a constant background $\phi_{0}$ gives the
effective coupling $G_{N,\mathrm{eff}}=e^{2\phi_{0}}G_{N}$ when lengths
are measured in asymptotically normalized string-frame coordinates. 

For a target space with boundary, $I_{0}$ includes the boundary terms
needed for a well-posed variational principle and, in an asymptotically
AdS setting, the corresponding counterterms. They must be transformed
consistently when integrating the dilaton Laplacian by parts. At fixed
asymptotic sources they do not alter the local conical entropy term.

The displayed heat-kernel derivation is bosonic. For the later type
IIB application we use the universal metric--dilaton term of the
type II tree-level NS--NS action, whose coefficient is likewise fixed
by superstring sphere amplitudes. Smooth NS--NS and RR matter terms
contain no explicit curvature coupling at two-derivative order, so
they do not modify the local area density. 

\subsection{Dilaton-weighted black hole entropy from off-shell effects}

We now apply this framework to the calculation of black hole entropy
by varying the conical deficit at the tip of the Euclidean cigar.
This requires an off-shell extension of the conventional worldsheet
formulation \cite{Callan:1985ia,Ahmadain:2022tew}. In the usual on-shell
formulation, the background fields are required to satisfy the conditions
of Weyl invariance,
\begin{equation}
\beta_{\mu\nu}^{g}=0,\qquad\beta^{\phi}=0,
\end{equation}
together with $\beta_{\mu\nu}^{B}=0$ when the Kalb--Ramond field
is included. Under these conditions, the worldsheet theory is a conformal
field theory and the target space background satisfies the corresponding
string equations of motion. The essential feature of the Tseytlin
prescription employed by Ahmadain and Wall is that the nonlinear sigma
model can also be defined away from a conformal fixed point. When
at least one beta function is nonzero, 
\begin{equation}
\bigl(\beta_{\mu\nu}^{g},\beta_{\mu\nu}^{B},\beta^{\phi}\bigr)\neq0,
\end{equation}
the worldsheet theory is no longer a CFT, but it remains a well-defined
two-dimensional quantum field theory once a worldsheet ultraviolet
cutoff $\epsilon$ and a Weyl frame have been specified. Consequently,
the regulated sphere functional $K_{0}\left(g,\phi;\epsilon\right)$
can be evaluated on backgrounds that do not satisfy the target-space
equations of motion. The connected genus-zero target-space functional
and the corresponding Euclidean action are defined by
\begin{equation}
Z_{0}=\frac{\partial K_{0}}{\partial\log\epsilon},\qquad I_{0}=-Z_{0}
\end{equation}
Schematically, if $\omega^{i}$ denotes the collection of worldsheet
couplings, including $g_{\mu\nu}$, $B_{\mu\nu}$, and $\phi$, the
variation of the action is related to the worldsheet beta functions
by 
\begin{equation}
E_{i}\coloneqq\frac{\partial I_{0}}{\partial\omega^{i}}=-\kappa_{ij}\beta^{j},
\end{equation}
where $\kappa_{ij}$ is a Zamolodchikov-type metric on the space of
couplings. Therefore,
\begin{equation}
\beta^{i}=0\qquad\Longleftrightarrow\qquad\frac{\delta I_{0}}{\delta\omega^{i}}=0,
\end{equation}
The term \textquotedblleft off shell\textquotedblright{} therefore
does not mean that this prescription produces different physical equations
of motion. Rather, it means that it defines the complete functional
$I_{0}\left[g,\phi,\ldots\right]$ on backgrounds away from its stationary
points. The usual on-shell backgrounds are recovered as extreme of
this functional, whereas the off-shell prescription also determines
its value when the beta functions are nonzero. This additional information
is essential for the conical calculation of black hole entropy, because
varying the Euclidean angular period away from its regular value produces
a conical background that does not satisfy the source-free classical
equations of motion.

In the notation of Ahmadain and Wall, the sphere contribution to the
black hole entropy is
\begin{align}
S_{\mathrm{BH}} & =\left.\left(1-\beta\partial_{\beta}\right)Z_{0}\left[\mathcal{M}_{\beta}\right]\right|_{\beta=2\pi}\nonumber \\
 & =\left.\left(1-\beta\partial_{\beta}\right)\frac{\partial K_{0}\left[\mathcal{M}_{\beta}\right]}{\partial\log\epsilon}\right|_{\beta=2\pi}\nonumber \\
 & =\left.\left(\beta\partial_{\beta}-1\right)I_{0}\left[\mathcal{M}_{\beta}\right]\right|_{\beta=2\pi}.
\end{align}
The parameter $\beta$ here is the dimensionless angular period. It
equals $2\pi$ on the regular cone and is distinct from the dimensionful
Hawking inverse temperature. Here $I_{0}\left[\mathcal{M}_{\beta}\right]$
denotes the value of the same sphere-derived functional $I_{0}\left[g,\phi,\ldots\right]$
on the conical background $\mathcal{M}_{\beta}$ , whereas $I_{0}[\mathcal{M}_{2\pi}]$
denotes its value on the smooth Euclidean black hole geometry. Our
immediate aim is therefore to determine the part of $I_{0}[\mathcal{M}_{\beta}]$
that is not simply proportional to the angular period $\beta$.

Near the bifurcation surface $\Gamma_{H}$ of a nonextremal stationary
horizon, the smooth Euclidean geometry has the local Rindler form
\begin{equation}
\mathrm{d}s^{2}=\mathrm{d}\rho^{2}+\rho^{2}\mathrm{d}\tau^{2}+\gamma_{ij}\left(Y\right)\mathrm{d}Y^{i}\mathrm{d}Y^{j}+O\left(\rho^{2}\right),\qquad\tau\sim\tau+2\pi,
\end{equation}
where $Y^{i}$ are coordinates on $\Gamma_{H}$. The surface $\rho=0$
is the tip of the Euclidean cigar, at which the $\tau$ circle shrinks
to zero size. The off-shell geometry $\mathcal{M}_{\beta}$ is obtained
by holding the local metric and matter fields fixed while changing
only the angular identification,
\begin{equation}
\tau\sim\tau+\beta.
\end{equation}
At proper radius $\rho$, the circumference of the angular circle
is then $\beta\rho$, rather than $2\pi\rho$. Thus, for $\beta\neq2\pi$,
the geometry has a conical defect of angle $2\pi-\beta$ at the fixed
surface $\Gamma_{H}$ \cite{Fursaev:1995ef}. Since the action Eq.
(\ref{eq:AW action}) is initially defined for smooth target space
fields, we regulate the conical tip by replacing it with a smooth
cap of radius $r_{*}$. The local metric is taken to be
\begin{equation}
\mathrm{d}s^{2}=\mathrm{d}\rho^{2}+f_{\beta}\left(\rho\right)^{2}\mathrm{d}\tau^{2}+\gamma_{ij}\left(Y\right)\mathrm{d}Y^{i}\mathrm{d}Y^{j}+O\left(\rho^{2}\right),\qquad0\leq\tau<\beta,
\end{equation}
where 
\begin{equation}
f_{\beta}\left(0\right)=0,\qquad f_{\beta}^{\prime}\left(0\right)=\frac{2\pi}{\beta},\qquad f_{\beta}\left(\rho\right)=\rho\quad\left(\rho\geq r_{*}\right).
\end{equation}
For the two-dimensional metric $g_{ab}$,
\begin{equation}
\mathrm{d}s_{\left(2\right)}^{2}=\mathrm{d}\rho^{2}+f_{\beta}\left(\rho\right)^{2}\mathrm{d}\tau^{2}.
\end{equation}
where $a,b$ are indices in the two-dimensional normal plane. Its
integrated curvature is
\begin{equation}
\int_{\mathcal{M}_{\beta}^{\left(2\right)}}\mathrm{d}^{2}X\sqrt{g^{\left(2\right)}}R^{\left(2\right)}=2\left(2\pi-\beta\right),
\end{equation}
which implies
\begin{equation}
\sqrt{g^{\left(2\right)}}R^{\left(2\right)}=2\left(2\pi-\beta\right)\delta^{\left(2\right)}\left(X\right)\label{eq:conical}
\end{equation}
The distributional formula denotes the singular part in the sharp-cap
limit. The conical response is computed to first order in $\beta-2\pi$
at fixed nonzero $r_{*}$ before taking $r_{*}\to0$. This order of
limits extracts the two-derivative entropy coefficient. This localized
curvature signals that $\mathcal{M}_{\beta}$ is off shell. For every
nonzero smoothing scale $r_{*}$, the background is smooth and the
usual local sigma model $\beta$ functions apply. With constant dilaton
and vanishing $H$ near the tip, 
\begin{equation}
\beta_{ab}^{g}=\alpha^{\prime}R_{ab}+O\left(\alpha^{\prime2}\right).
\end{equation}
In two dimensions $R_{ab}^{\left(2\right)}=\frac{1}{2}g_{ab}^{\left(2\right)}R^{\left(2\right)}$.
Combining this identity with Eq. (\ref{eq:conical}) and then taking
$r_{*}\to0$ gives 
\begin{equation}
\left.\beta_{ab}^{g}\right|_{\mathrm{tip}}=\alpha^{\prime}\left(2\pi-\beta\right)g_{ab}^{\left(2\right)}\frac{\delta^{\left(2\right)}\left(X\right)}{\sqrt{g^{\left(2\right)}}}+O\left(\left(\beta-2\pi\right)^{2},\alpha^{\prime2}\right).
\end{equation}
This delta function is a distribution in the target space, and the
worldsheet itself remains a single smooth sphere. Smooth background
fluxes and a smooth dilaton contribute regular terms to the beta functions
and do not cancel this singular curvature contribution. Based on the
gradient relation introduced above,
\begin{equation}
E_{i}=-\kappa_{ij}\beta^{j},
\end{equation}
this localized beta function is equivalent to a delta-function contribution
to the target-space equation-of-motion:
\begin{equation}
\left.E_{i}\right|_{\mathrm{tip}}\neq0,\qquad\beta\neq2\pi.
\end{equation}
In the sphere-amplitude description, the same nonzero effect appears
as a localized graviton one-point function, or graviton tadpole
\begin{equation}
A_{0,1}^{g}=\frac{\partial}{\partial\log\epsilon}K_{0,1}^{g}\sim-\int_{\mathcal{M}_{\beta}}\mathrm{d}^{D}X\sqrt{g}\delta g^{ab}\left(X\right)E_{\left(g\right)ab}\left(X\right),
\end{equation}

We now return to the action (\ref{eq:AW action}) and evaluate the
corresponding black hole entropy. Near the horizon of this metric,
the measure factorizes as $\sqrt{g}=\sqrt{g^{\left(2\right)}}\sqrt{\gamma}+O\left(\rho\right)^{2}$,
while smooth replica symmetry implies 
\begin{equation}
\phi\left(\rho,\tau,Y\right)=\phi_{H}\left(Y\right)+O\left(\rho\right)^{2},
\end{equation}
where $\phi_{H}\left(Y\right)\equiv\phi|_{\Gamma_{H}}$. Then, the
curvature of the action (\ref{eq:AW action}) localized in the cap
gives 
\begin{align}
\left.\int_{\mathcal{M}_{\beta}}\mathrm{d}^{D}X\sqrt{g}e^{-2\phi}R\right|_{\mathrm{cap}} & =\left(\int_{\mathcal{M}_{\beta}^{\left(2\right)}}\mathrm{d}^{2}X\sqrt{g^{\left(2\right)}}R^{\left(2\right)}\right)\left(\int_{\Gamma_{H}}\mathrm{d}^{D-2}Y\sqrt{\gamma}e^{-2\phi_{H}}\right)\nonumber \\
 & =2\left(2\pi-\beta\right)\mathcal{A}_{\phi}\left(\Gamma_{H}\right)+O\left(r_{*}^{2}\right)+O\left(\left(\beta-2\pi\right)^{2}\right),\label{eq:AW tip}
\end{align}
where 
\begin{equation}
\mathcal{A}_{\phi}\left(\Gamma_{H}\right):=\int_{\Gamma_{H}}\mathrm{d}^{D-2}Y\sqrt{\gamma}e^{-2\phi_{H}},\qquad\sqrt{g^{\left(2\right)}}R^{\left(2\right)}=2\left(2\pi-\beta\right)\delta^{\left(2\right)}\left(X\right).
\end{equation}
Moreover, since $\nabla^{2}\phi$ is bounded for the smooth $\phi_{H}\left(Y\right)$,
the kinematic dilaton term of the action (\ref{eq:AW action}) becomes
\begin{equation}
\left.\int_{\mathcal{M}_{\beta}}\mathrm{d}^{D}X\sqrt{g}e^{-2\phi}\nabla^{2}\phi\right|_{\mathrm{cap}}=O\left(r_{*}^{2}\right),\qquad\mathrm{as}\quad r_{*}\rightarrow0.
\end{equation}
Thus the dilaton enters the leading tip term only through the weight
$e^{-2\phi_{H}}$ in $\mathcal{A}_{\phi}$.

Away from the cap, changing the range of the $\tau$ integral from
$2\pi$ to $\beta$ simply multiplies every regular bulk and asymptotic
boundary contribution by $\beta/\left(2\pi\right)$. Substituting
Eq. (\ref{eq:AW tip}) into the action (\ref{eq:AW action}) gives
\begin{equation}
I_{0}\left[\mathcal{M}_{\beta}\right]=\frac{\beta}{2\pi}I_{0}\left[\mathcal{M}_{2\pi}\right]+I_{\mathrm{tip}}\left(\beta\right)+O\left(\left(\beta-2\pi\right)^{2},r_{*}^{2},\alpha^{\prime}\right),
\end{equation}
where
\begin{equation}
I_{\mathrm{tip}}\left(\beta\right)=-\frac{1}{16\pi G_{N}}\left[2\left(2\pi-\beta\right)\mathcal{A}_{\phi}\left(\Gamma_{H}\right)\right]=\frac{\beta-2\pi}{8\pi G_{N}}\mathcal{A}_{\phi}\left(\Gamma_{H}\right).
\end{equation}
Equivalently, 
\begin{equation}
I_{0}\left[\mathcal{M}_{\beta}\right]-\frac{\beta}{2\pi}I_{0}\left[\mathcal{M}_{2\pi}\right]=\frac{\beta-2\pi}{8\pi G_{N}}\mathcal{A}_{\phi}\left(\Gamma_{H}\right)+O\left(\left(\beta-2\pi\right)^{2},r_{*}^{2},\alpha^{\prime}\right),
\end{equation}
Since $-I_{0}=Z_{0}$, utilizing sphere contribution to the black
hole entropy proposed by Ahmadain and Wall, we have
\begin{align}
S_{\mathrm{BH}} & =\left.\left(1-\beta\partial_{\beta}\right)Z_{0}\left[\mathcal{M}_{\beta}\right]\right|_{\beta=2\pi}\nonumber \\
 & =\left.\left(\beta\partial_{\beta}-1\right)I_{0}\left[\mathcal{M}_{\beta}\right]\right|_{\beta=2\pi}.\nonumber \\
 & =\frac{\mathcal{A}_{\phi}\left(\Gamma_{H}\right)}{4G_{N}}.
\end{align}
For the constant-dilaton Rindler background, $\mathcal{A}_{\phi}=A_{\perp}$,
the entropy reduces to the Ahmadain and Wall's result $S_{\mathrm{BH}}=A_{\perp}/4G_{N}$.
At two-derivative order, the same dilaton-weighted area follows from
the Noether-charge entropy of the string-frame action \cite{Wald:1993nt,Iyer:1994ys}.

\section{The stringy origin of the norm bound}

\label{sec:norm-bound}

The calculation in the previous section already completes the genus-zero
entropy derivation for a prescribed stationary Killing horizon. Its
result can be written as
\begin{equation}
S_{\mathrm{BH}}=\frac{\mathcal{A}_{\phi}\left(\Gamma_{H}\right)}{4G_{N}},\qquad\mathcal{A}_{\phi}\left(\Gamma_{H}\right):=\int_{\Gamma_{H}}e^{-2\phi_{H}}dA_{S},
\end{equation}
where $dA_{S}=\sqrt{\gamma}\mathrm{d}^{D-2}Y$. The bifurcation surface
$\Gamma_{H}$ is fixed in advance by the stationary black hole geometry.
In the Euclidean continuation, it is the locus at which the thermal
circle shrinks to zero size. The calculation above therefore evaluates
the entropy on a given surface. It does not search over different
codimension-two surfaces.

Because the genus-zero two-derivative target-space action is a local
Einstein--dilaton functional, the change in the action caused by
a small conical deficit is determined entirely by the fields near
the codimension-two defect. The same regulated-cone calculation can
therefore be applied to any smooth codimension-two candidate surface
$\Gamma$. It assigns to $\Gamma$ the local entropy functional
\begin{equation}
S_{{\rm loc}}^{\left(0\right)}\left[\Gamma\right]=\frac{1}{4G_{N}}\int_{\Gamma}e^{-2\phi}dA_{S}=\int_{\Gamma}\rho_{S}\left(x\right)dA_{S},\qquad\rho_{S}\left(x\right):=\frac{e^{-2\phi\left(x\right)}}{4G_{N}}.\label{eq:local entropy}
\end{equation}
Thus $\rho_{S}\left(x\right)$ is the entropy cost per unit string-frame
area assigned by the sphere-derived target-space action. Equivalently,
\begin{equation}
\rho_{S}\left(x\right)=\frac{\delta S_{{\rm loc}}^{\left(0\right)}}{\delta A_{S}\left(x\right)}.
\end{equation}
Evaluating this functional on the prescribed bifurcation surface,
$\Gamma=\Gamma_{H}$, reproduces the black hole entropy:
\begin{equation}
S_{{\rm BH}}^{\left(0\right)}=S_{{\rm loc}}^{\left(0\right)}\left[\Gamma_{H}\right]=\int_{\Gamma_{H}}\rho_{S}\left(x\right)dA_{S}.
\end{equation}
This equality gives an evaluation of $S_{\mathrm{loc}}^{(0)}$ on
a surface whose location is already fixed by the Killing geometry. 

But for the generic boundary region $A$, how does the local functional
$S_{{\rm loc}}^{\left(0\right)}\left[\Gamma\right]$ determine the
entropy? Let $A$ be a spatial region on the boundary of a static
bulk time slice. There are generally many bulk codimension-two surfaces
$\Gamma$ satisfying the anchoring condition $\partial\Gamma=\partial A$.
The admissible surfaces must also satisfy the homology condition $\Gamma\sim A$.
The fixed-cone calculation assigns a value $S_{{\rm loc}}^{\left(0\right)}\left[\Gamma\right]$
to each such candidate surface, but it does not determine which candidate
is selected by the boundary density matrix:
\begin{equation}
S_{{\rm loc}}^{\left(0\right)}\left[\Gamma\right]\quad\not\Rightarrow\quad S_{A}^{\left(0\right)}=\min_{\Gamma\sim A}S_{{\rm loc}}^{\left(0\right)}\left[\Gamma\right],
\end{equation}
without an additional surface-selection argument. In other words,
it cannot, by itself, provide a minimization principle over arbitrary
codimension-two surfaces. Therefore, the purpose of the following
subsection is to derive this selection rule from the gravitational
replica construction. In other words, the Ahmadain--Wall off-shell
calculation determines the entropy density assigned to each candidate
surface, while the replica construction selects the corresponding
weighted minimal cut. Then, in the following subsections, we will
show that the convex duality converts this weighted minimal cut into
a divergenceless flow obeying a norm bound which provides the string
derived norm bound. Completing this chain relates the sphere derived
entropy functional to the bit-thread formulation and provides the
norm bound required for the subsequent identification of the bit threads
with the appropriately string charge density.

\subsection{From the replica saddle to the weighted minimal cut}

We now combine the entropy density
\begin{equation}
\rho_{S}\left(x\right)=\frac{e^{-2\phi\left(x\right)}}{4G_{N}},
\end{equation}
obtained from the Ahmadain--Wall sphere calculation with the standard
Lewkowycz--Maldacena gravitational replica argument \cite{Lewkowycz:2013nqa}.
\emph{This combination requires an additional holographic step}. The
fixed-cone calculation above does not, by itself, introduce a boundary
density matrix or its replica partition functions. To pass from the
entropy of a prescribed Killing horizon to that of a generic boundary
region, we assume that the target-space string background admits a
semiclassical holographic description. We also assume that replica
symmetry is unbroken in the bulk and that the dominant replica saddles
admit an analytic continuation to a neighborhood of $n=1$. In this
subsection, $n$ denotes the replica index but not the number of integrated
worldsheet insertions used in the perturbative expansion of $K_{0}$
previously.

The connection between the worldsheet calculation and the gravitational
replica construction comes from the same tree-level target space action
$I_{0}$, including the boundary terms appropriate to the chosen asymptotic
sources. In the notation introduced above 
\begin{equation}
Z_{0}\left[\Psi\right]=\frac{\partial K_{0}\left[\Psi;\epsilon\right]}{\partial\log\epsilon}=-I_{0}\left[\Psi\right],
\end{equation}
where $\Psi=\left\{ g_{\mu\nu},B_{\mu\nu},\phi,\ldots\right\} $ collectively
denotes the target space background fields as we introduced before.
Note that $Z_{0}\left[\Psi\right]$ here is the connected genus-zero
sphere contribution evaluated in the background $\Psi$. Now, Let
$\mathcal{B}$ denote a choice of asymptotic target space boundary
data, and let $\Psi_{\mathrm{cl}}\left(\mathcal{B}\right)$ be the
corresponding classical target space saddle. Summing over any number
$l$ of disconnected sphere components give the connected sphere contribution:

\begin{equation}
\mathcal{Z}_{\mathrm{bulk}}^{\left(0\right)}\left[\mathcal{B}\right]:=\sum_{l=0}^{\infty}\frac{1}{l!}\left(Z_{0}\left[\Psi_{\mathrm{cl}}\left(\mathcal{B}\right)\right]\right)^{l}=\exp\left(Z_{0}\left[\Psi_{\mathrm{cl}}\left(\mathcal{B}\right)\right]\right)=\exp\left(-I_{0}\left[\Psi_{\mathrm{cl}}\left(\mathcal{B}\right)\right]\right).
\end{equation}
where $\mathcal{Z}_{\mathrm{bulk}}^{\left(0\right)}\left[\mathcal{B}\right]$
denotes the genus-zero bulk string on-shell partition function with
asymptotic boundary data $\mathcal{B}$. Next, let $\mathcal{B}_{n}$
denote the $n$-fold replica boundary geometry associated with the
boundary region $A$. The boundary replica construction defines
\begin{equation}
Z\left(n\right)\coloneqq\mathrm{Tr}\rho_{A}^{n},
\end{equation}
where $\rho_{A}$ may be unnormalized. Holographic duality identifies
this boundary quantity with the bulk string partition function subject
to the replica boundary conditions $\mathcal{B}_{n}$. In the genus-zero
semiclassical approximation, the corresponding bulk path integral
is dominated by the smooth on-shell replica saddle $\mathcal{M}_{n}$.
Consequently,

\begin{equation}
Z\left(n\right)=Z_{\mathrm{boundary}}\left[\mathcal{B}_{n}\right]\simeq\mathcal{Z}_{\mathrm{bulk}}^{\left(0\right)}\left[\mathcal{B}_{n}\right]=\exp\left(Z_{0}\left[\mathcal{M}_{n}\right]\right)=\exp\left(-I_{0}\left[\mathcal{M}_{n}\right]\right).
\end{equation}
Equivalently,
\begin{equation}
\log Z\left(n\right)=Z_{0}\left[\mathcal{M}_{n}\right]=-I_{0}\left[\mathcal{M}_{n}\right].
\end{equation}
In particular,
\begin{equation}
Z\left(1\right)\simeq\exp\left(Z_{0}\left[\mathcal{M}_{1}\right]\right)=\exp\left(-I_{0}\left[\mathcal{M}_{1}\right]\right).
\end{equation}
Note that $Z_{0}\neq Z\left(1\right)$, where $Z_{0}$ is the connected
sphere contribution, whereas $Z\left(1\right)$ is the full boundary
partition function at replica number $n=1$. At genus zero, they are
related by exponentiation above. The normalized reduced density matrix
is
\begin{equation}
\widehat{\rho}_{A}\coloneqq\frac{\rho_{A}}{\mathrm{Tr}\rho_{A}}=\frac{\rho_{A}}{Z\left(1\right)},
\end{equation}
and hence
\begin{equation}
\mathrm{Tr}\widehat{\rho}_{A}^{n}=\frac{Z\left(n\right)}{Z\left(1\right)^{n}}.
\end{equation}
The von Neumann entropy is therefore
\begin{align}
S_{A}^{\left(0\right)} & =-\left.n\partial_{n}\left[\log Z\left(n\right)-n\log Z\left(1\right)\right]\right|_{n=1}\nonumber \\
 & =\left.n\partial_{n}\left(I_{0}\left[\mathcal{M}_{n}\right]-nI_{0}\left[\mathcal{M}_{1}\right]\right)\right|_{n=1}\nonumber \\
 & =\left.\left(n\partial_{n}-1\right)I_{0}\left[\mathcal{M}_{n}\right]\right|_{n=1}.
\end{align}
Moreover, it is worth noting that $\mathcal{M}_{n}$ here is the smooth
on-shell Lewkowycz--Maldacena replica saddle and must not be confused
with the Ahmadain--Wall off-shell cone $\mathcal{M}_{\beta}$.

For integer $n>1$, the bulk saddle $\mathcal{M}_{n}$ is assumed
to preserve the $\mathbb{Z}_{n}$ replica symmetry. It contains a
special codimension-two fixed locus at which the replica circle shrinks.
Following Lewkowycz and Maldacena \cite{Lewkowycz:2013nqa}, we assume
that the family of replica saddles admits an analytic continuation
toward $n=1$, and we denote by $\Gamma\subset\mathcal{M}_{1}$ the
limiting position of this special surface in the $n=1$ geometry.
At this stage, the location of $\Gamma$ is not yet known. We will
come back to determine it soon. Although the complete $n$-dependent
solution $\mathcal{M}_{n}$ is generally not known, only its first-order
variation around $n=1$ is required. We therefore introduce a smooth
auxiliary geometry $\widetilde{\mathcal{M}}_{n}\left(r_{*}\right)$.
Outside a cap neighborhood of radius $r_{*}$ around the undetermined
surface $\Gamma$, it is constructed from $n$ cyclically glued copies
of the $n=1$ geometry. Inside the cap, the conical tip is replaced
by a smooth regulated squashed cone \cite{Fursaev:2013fta,Dong:2016fnf}.
Therefore, $\widetilde{\mathcal{M}}_{n}$ is off-shell and relates
to the Ahmadain--Wall construction through the same leading normal-plane
opening angle $\beta=2\pi n$. For a generic surface with nonzero
extrinsic curvature, the regulated squashed cone is not identical
to a product Killing-horizon cone. Thus $\mathcal{M}_{n}$, $\widetilde{\mathcal{M}}_{n}$,
and the fixed-field cone are distinct away from $n=1$, while they
reduce to the same background $\mathcal{M}_{1}$ at $n=1$. Therefore,
the two calculations play complementary roles. The Ahmadain--Wall
calculation determines the local conical-defect contribution contained
in $I_{0}$, and hence fixes the local entropy density $\rho_{S}\left(x\right)$.
The Lewkowycz--Maldacena replica construction determines which codimension-two
locus is selected by the boundary density matrix.

To calculate the entropy $S_{A}^{\left(0\right)}$, let us define
the corresponding off-shell partition function by
\begin{equation}
\log Z^{\mathrm{off}}\left(n\right)\coloneqq-I_{0}\left[\widetilde{\mathcal{M}}_{n}\left(r_{*}\right)\right].
\end{equation}
The auxiliary configuration has the same asymptotic sources as $\mathcal{M}_{n}$
and may be chosen so that, after identifying the two field configurations
in a fixed gauge, 
\begin{equation}
\delta\Psi_{n}:=\widetilde{\Psi}_{n}-\Psi_{n}=O\left(n-1\right),
\end{equation}
where $\Psi$ denotes collectively the metric, dilaton, and any other
target fields as before. The limit $n\to1$ is taken before the sharp
cap limit $r_{*}\to0$.

Based on $S_{A}^{\left(0\right)}=-\left.n\partial_{n}\left[\log Z\left(n\right)-n\log Z\left(1\right)\right]\right|_{n=1}$,
adding and subtracting the off-shell partition function give
\begin{align}
\log Z\left(n\right)-n\log Z\left(1\right)= & \left(\log Z\left(n\right)-\log Z^{\mathrm{off}}\left(n\right)\right)\nonumber \\
 & +\left(\log Z^{\mathrm{off}}\left(n\right)-n\log Z\left(1\right)\right).\label{eq:replica add}
\end{align}
Expanding the auxiliary configuration around the true on-shell saddle
gives
\begin{align}
I_{0}\left[\widetilde{\Psi}_{n}\right]-I_{0}\left[\Psi_{n}\right]= & \int_{\mathcal{M}_{n}}E_{\Psi}\left(\Psi_{n}\right)\delta\Psi_{n}\nonumber \\
 & +\int_{\partial\mathcal{M}_{n}}\Theta\left(\Psi_{n},\delta\Psi_{n}\right)+O\left(\delta\Psi_{n}^{2}\right).\label{eq:variation saddle}
\end{align}
The bulk term vanishes because $\mathcal{M}_{n}$ is a saddle, $E_{\Psi}\left(\Psi_{n}\right)=0$.
The asymptotic term vanishes because the sources are held fixed and
the complete action has a well-posed variational principle. Since
the cap has been filled in smoothly, there is no internal boundary
in Eq. (\ref{eq:variation saddle}). It follows that
\begin{equation}
\log Z\left(n\right)-\log Z^{\mathrm{off}}\left(n\right)=I_{0}\left[\widetilde{\mathcal{M}}_{n}\right]-I_{0}\left[\mathcal{M}_{n}\right]=O\left(\left(n-1\right)^{2}\right).\label{eq:first bracket}
\end{equation}
For the second bracket in Eq. (\ref{eq:replica add}), setting $\beta=2\pi n$
in the fixed-cone result gives
\begin{align}
\log Z^{\mathrm{off}}\left(n\right)-n\log Z\left(1\right)= & -\frac{2\pi\left(n-1\right)}{8\pi G_{N}}\int_{\Gamma}e^{-2\phi}\mathrm{d}A_{S}\nonumber \\
 & +O\left(\left(n-1\right)^{2}\right)+\left(n-1\right)O\left(\alpha^{\prime}\right)\nonumber \\
= & -\left(n-1\right)S_{{\rm loc}}^{\left(0\right)}\left[\Gamma\right]+O\left(\left(n-1\right)^{2}\right)+\left(n-1\right)O\left(\alpha^{\prime}\right),\label{eq:second bracket}
\end{align}
where we use the important result
\begin{equation}
I_{0}[\widetilde{\mathcal{M}}_{n}]-nI_{0}[\mathcal{M}_{1}]=I_{\mathrm{tip}}\left(n\right)+O\left(\left(n-1\right)^{2}\right).
\end{equation}
Combining Eqs. (\ref{eq:replica add}), (\ref{eq:first bracket}),
and (\ref{eq:second bracket}) gives 
\begin{equation}
\log Z\left(n\right)-n\log Z\left(1\right)=-\left(n-1\right)S_{{\rm loc}}^{\left(0\right)}\left[\Gamma\right]+O\left(\left(n-1\right)^{2}\right)+\left(n-1\right)O\left(\alpha^{\prime}\right).
\end{equation}
Using $S_{A}^{\left(0\right)}=-\left.n\partial_{n}\left[\log Z\left(n\right)-n\log Z\left(1\right)\right]\right|_{n=1}$,
we have
\begin{equation}
S_{A}^{\left(0\right)}=S_{{\rm loc}}^{\left(0\right)}\left[\Gamma\right]=\frac{1}{4G_{N}}\int_{\Gamma}e^{-2\phi}dA_{S},
\end{equation}
up to higher-derivative corrections. 

Finally, it remains to determine the location of $\Gamma$, which
must be compatible with replica regularity. Near this surface, introduce
complex coordinates $z,\bar{z}$ in the two-dimensional normal plane.
Writing $n=1+\varepsilon$, the linearized Einstein-frame Ricci tensor
contains the potentially singular terms \cite{Lewkowycz:2013nqa}:
\begin{align}
\delta R_{zz}^{E} & =-\varepsilon\frac{K_{z}^{E}}{z}-\frac{1}{2}\partial_{z}^{2}\delta\gamma+\cdots,\nonumber \\
\delta R_{\bar{z}\bar{z}}^{E} & =-\varepsilon\frac{K_{\bar{z}}^{E}}{\bar{z}}-\frac{1}{2}\partial_{\bar{z}}^{2}\delta\gamma+\cdots,
\end{align}
where $\delta\gamma=\gamma_{E}^{ij}\delta g_{ij}^{E}$, $K_{z}^{E}=\frac{1}{2}\left(K_{1}^{E}-iK_{2}^{E}\right)$.
It also can be written as
\begin{equation}
R_{zz}^{E}\supset-\varepsilon\frac{K_{z}^{E}}{z},\qquad R_{\bar{z}\bar{z}}^{E}\supset-\varepsilon\frac{K_{\bar{z}}^{E}}{\bar{z}}.
\end{equation}
display the geometrically important poles, but they are not by themselves
the complete derivation. The perturbation $\delta\gamma$ might apparently
cancel one of these poles. Lewkowycz and Maldacena show, however,
that no single-valued $2\pi$-periodic perturbation can cancel both
the $zz$ and $\bar{z}\bar{z}$ poles unless
\begin{equation}
K_{z}^{E}=K_{\bar{z}}^{E}=0,
\end{equation}
Thus replica regularity selects an extremal surface in the Einstein-frame
geometry.

The same condition can be expressed directly in string-frame variables
by varying Eq. (\ref{eq:local entropy}). For a normal displacement
$\delta X^{\mu}=\delta X^{a}n_{a}^{\mu}$, and with the convention
$\delta\left(\mathrm{d}A_{S}\right)=K_{a}\delta X^{a}\mathrm{d}A_{S}$,
\begin{equation}
\delta_{\Gamma}\int_{\Gamma}e^{-2\phi}\mathrm{d}A_{S}=\int_{\Gamma}e^{-2\phi}\left(K_{a}-2n_{a}^{\mu}\nabla_{\mu}\phi\right)\delta X^{a}\mathrm{d}A_{S}.
\end{equation}
Hence the string-frame replica-extremality equation is
\begin{equation}
K_{a}-2n_{a}^{\mu}\nabla_{\mu}\phi=0.
\end{equation}
The boundary replica construction also supplies the anchoring and
homology conditions
\begin{equation}
\partial\Gamma=\partial A,\qquad\Gamma\sim A.
\end{equation}

Now suppose that the bulk geometry is static and invariant under reflection
about a time-symmetric Riemannian slice. The surface-selection problem
can then be formulated entirely on this slice. The replica boundary-value
problem may admit several smooth replica-symmetric saddle branches
\begin{equation}
\mathcal{M}_{n}^{\left(1\right)},\qquad\mathcal{M}_{n}^{\left(2\right)},\qquad\ldots,
\end{equation}
with the same asymptotic replica boundary data $\mathcal{B}_{n}$.
Here $i$ of $\mathcal{M}_{n}^{\left(i\right)}$ labels distinct semiclassical
saddle branches. We consider branches that reduce to the same dominant
background $\mathcal{M}_{1}$ as $n\to1$, although the codimension-two
loci at which their replica circles close may approach different surfaces,
\begin{equation}
\Gamma_{n}^{\left(i\right)}\longrightarrow m_{i}\qquad\left(n\to1\right).
\end{equation}
Each $m_{i}$ is an admissible weighted-extremal cut satisfying
\begin{equation}
\partial m_{i}=\partial A,\qquad m_{i}\sim A,\qquad K_{a}-2n_{a}^{\mu}\nabla_{\mu}\phi=0,
\end{equation}
where $A$ is the boundary region. Thus, the extremality equation
derived above determines the local candidate surfaces. Applying the
local conical-defect calculation to each saddle branch gives
\begin{equation}
I_{0}\left[\mathcal{M}_{n}^{\left(i\right)}\right]=nI_{0}\left[\mathcal{M}_{1}\right]+\left(n-1\right)S_{{\rm loc}}^{\left(0\right)}\left[m_{i}\right]+O\left(\left(n-1\right)^{2}\right)+\left(n-1\right)O\left(\alpha^{\prime}\right).
\end{equation}
where
\begin{equation}
S_{{\rm loc}}^{\left(0\right)}\left[m_{i}\right]=\frac{1}{4G_{N}}\int_{m_{i}}e^{-2\phi}dA_{S}
\end{equation}
In the semiclassical approximation, the replica partition function
is a sum over these saddle contributions:
\begin{align}
\begin{aligned}Z\left(n\right) & \simeq\sum_{i}\exp\left(-I_{0}\left[\mathcal{M}_{n}^{\left(i\right)}\right]\right)\\
 & =e^{-nI_{0}\left[\mathcal{M}_{1}\right]}\sum_{i}\exp\left[-\left(n-1\right)S_{{\rm loc}}^{\left(0\right)}\left[m_{i}\right]+O\left(\left(n-1\right)^{2}\right)+\left(n-1\right)O\left(\alpha^{\prime}\right)\right].
\end{aligned}
\end{align}
For $n>1$, the factor $n-1$ is positive. Since $S_{{\rm loc}}^{\left(0\right)}\left[m_{i}\right]=O\left(G_{N}^{-1}\right)$,
the saddle with the smallest $S_{{\rm loc}}^{\left(0\right)}\left[m_{i}\right]$
is exponentially dominant in the classical limit. More explicitly,
for two branches $i$ and $j$,
\begin{equation}
\frac{\exp\left[-I_{0}\left[\mathcal{M}_{n}^{\left(j\right)}\right]\right]}{\exp\left[-I_{0}\left[\mathcal{M}_{n}^{\left(i\right)}\right]\right]}=\exp\left[-\left(n-1\right)\left(S_{{\rm loc}}^{\left(0\right)}\left[m_{j}\right]-S_{{\rm loc}}^{\left(0\right)}\left[m_{i}\right]\right)+\cdots\right].
\end{equation}
Hence, if 
\begin{equation}
S_{{\rm loc}}^{\left(0\right)}\left[m_{i}\right]<S_{{\rm loc}}^{\left(0\right)}\left[m_{j}\right],
\end{equation}
the $j$-th branch is exponentially suppressed relative to the $i$-th
branch. Taking the semiclassical limit before the continuation $n\to1^{+}$,
one obtains
\begin{equation}
\log Z\left(n\right)=-nI_{0}\left[\mathcal{M}_{1}\right]-\left(n-1\right)\;\min_{i}S_{{\rm loc}}^{\left(0\right)}\left[m_{i}\right]+O\left(\left(n-1\right)^{2}\right)+\left(n-1\right)O\left(\alpha^{\prime}\right).
\end{equation}
Since $\log Z\left(1\right)=-I_{0}\left[\mathcal{M}_{1}\right]$,
the replica formula gives 
\begin{align}
S_{A}^{\left(0\right)} & =-\left.n\partial_{n}\left[\log Z\left(n\right)-n\log Z\left(1\right)\right]\right|_{n=1}\nonumber \\
 & =\min_{i}\frac{1}{4G_{N}}\int_{m_{i}}e^{-2\phi}dA_{S}.
\end{align}
Therefore, assuming that all relevant dominant extremal branches admit
the required replica continuation, the minimization over saddle branches
may equivalently be written as a minimization over all admissible
cuts:
\begin{equation}
S_{A}^{\left(0\right)}=\min_{\substack{\partial m=\partial A\\
m\sim A
}
}\int_{m}\rho_{S}\left(x\right)dA_{S},\qquad\rho_{S}\left(x\right)=\frac{e^{-2\phi\left(x\right)}}{4G_{N}},
\end{equation}
at leading order in both the genus and derivative expansions. Here
$m\sim A$ denotes the homology condition. Thus, replica regularity
gives the weighted-extremality equation, whereas semiclassical saddle
dominance promotes the relevant extremum to the global weighted minimum.

\subsection{Convex optimization and the origin of the string-frame norm bound}

We now translate the weighted minimal cut obtained above into a flow
formulation. We work first at a finite radial cutoff with regulated
entangling anchors, so that the geometric optimization problem is
well defined. The cutoff may be removed after the usual entropy regularization.
To avoid confusion between the worldsheet and the bulk spatial slice,
we denote the genus-zero worldsheet by $\Sigma_{\mathrm{ws}}\simeq S^{2}$,
whereas $\Sigma$ denotes a static, time reflection symmetric bulk
slice. The induced string-frame metric on $\Sigma$ is denoted by
$h_{ij}$. A codimension-two spacetime surface becomes a codimension-one
cut $m\subset\Sigma$. All calculations in this subsection are performed
using the string-frame metric.

The origin of the norm bound can already be seen on an infinitesimal
surface element. Let $n^{i}$ be its unit normal, the infinitesimal
flux of a vector field $v^{i}$ is

\begin{equation}
d\Phi\left[v^{i}\right]=v^{i}n_{i}dA_{S},
\end{equation}
whereas the entropy capacity obtained from the sphere-derived functional
previously is

\begin{equation}
dS_{\mathrm{cap}}=\rho_{S}\left(x\right)dA_{S},\qquad\rho_{S}\left(x\right)=\frac{e^{-2\phi\left(x\right)}}{4G_{N}}.
\end{equation}
Demanding $d\Phi\left[v^{i}\right]\leq dS_{\mathrm{cap}}$ for every
orientation $n^{i}$ gives 

\begin{equation}
\left|v^{i}n_{i}\right|\leq\rho_{S}\left(x\right)\quad\text{for every unit }n^{i},
\end{equation}
which is equivalent to

\begin{equation}
|v|\leq\rho_{S}\left(x\right)=\frac{e^{-2\phi\left(x\right)}}{4G_{N}}.
\end{equation}
For completeness, we now derive the weighted Riemannian max-flow/min-cut
relation by following the convex-duality argument of ref. \cite{Headrick:2017ucz}.
The only modification is that the unit capacity used there is replaced
by the positive position-dependent capacity $\rho_{S}\left(x\right)$.
Correspondingly, the ordinary area and total-variation functionals
are replaced by their dilaton weighted counterparts.

Let $\Sigma$ be the static bulk slice equipped with the string-frame
metric $h_{ij}$. We allow its boundary to contain, in addition to
$A$ and $\bar{A}$, a component $R$ through which no flux is permitted.
Now, our aim is to show that

\begin{equation}
\sup_{v}\Phi\left[A\right],\qquad\Phi\left[A\right]:=\int_{A}v^{i}n_{i}dA_{S},
\end{equation}
subject to 
\begin{equation}
\nabla_{i}v^{i}=0,\qquad|v|\leq\rho_{S}\left(x\right),\qquad\left.v^{i}n_{i}\right|_{R}=0.
\end{equation}
all norms and volume elements are defined using the string-frame metric.
We assume that $\rho_{S}\left(x\right)$ is smooth and strictly positive.
The configuration $v^{i}=0$ then strictly satisfies the inequality
constraint. With the standard Riemannian flow--cut theorem, this
gives the strict feasibility used in the convex-duality argument \cite{Headrick:2017ucz,Boyd:2004fnq}.

To dualize the program, we introduce a scalar Lagrange multiplier
$\psi$ for the divergenceless constraint and a non-negative scalar
multiplier $\lambda$ for the norm bound based on ref. \cite{Headrick:2017ucz}.
Keeping the no-flux condition on $R$ as an implicit constraint, the
Lagrangian is

\begin{equation}
L\left[v,\psi,\lambda\right]=\int_{A}v^{i}n_{i}dA_{S}+\int_{\Sigma}\left[-\psi\nabla_{i}v^{i}+\lambda\bigl(\rho_{S}-|v|\bigr)\right]dV_{S}.
\end{equation}
Integrating the first bulk term by parts gives 

\begin{equation}
L\left[v,\psi,\lambda\right]=\int_{\partial\Sigma\setminus R}v^{i}n_{i}\left(\chi_{A}-\psi\right)dA_{S}+\int_{\Sigma}\left[v^{i}\nabla_{i}\psi-\lambda|v|+\lambda\rho_{S}\right]dV_{S}.
\end{equation}
where $\chi_{A}$ is the characteristic function of $A$ on $\partial\Sigma\setminus R$, 

\begin{equation}
\chi_{A}=\begin{cases}
1, & \text{on }A,\\
0, & \text{on }\bar{A}.
\end{cases}
\end{equation}
Note that $\partial\Sigma=A\cup\bar{A}\cup R$ and $\bar{A}:=\left(\partial\Sigma\setminus R\right)\setminus A$.
The maximization over the boundary values of $v^{i}$ is finite only
if

\begin{equation}
\left.\psi\right|_{A}=1,\qquad\left.\psi\right|_{\bar{A}}=0.
\end{equation}
In the interior, using Cauchy--Schwarz inequality $v^{i}\nabla_{i}\psi\leq|v||\nabla\psi|$,
the terms depending on $v^{i}$ obey 
\begin{equation}
v^{i}\nabla_{i}\psi-\lambda|v|\leq\left(|\nabla\psi|-\lambda\right)|v|.
\end{equation}
The pointwise supremum over $v^{i}$ is therefore finite if and only
if

\begin{equation}
\lambda\geq|\nabla\psi|,
\end{equation}
in which case the supremum of the $v^{i}$-dependent terms vanishes.
Strong convex duality then gives the dual program 
\begin{equation}
\inf_{\substack{\psi|_{A}=1,\ \psi|_{\bar{A}}=0\\
\lambda\geq|\nabla\psi|
}
}\int_{\Sigma}\rho_{S}\left(x\right)\lambda dV_{S}.
\end{equation}
Because $\rho_{S}\left(x\right)>0$, the minimum over $\lambda$ is
attained pointwise at
\begin{equation}
\lambda=|\nabla\psi|.
\end{equation}
 The dual problem therefore reduces to 
\begin{equation}
\inf_{\substack{\psi|_{A}=1,\ \psi|_{\bar{A}}=0}
}\int_{\Sigma}\rho_{S}\left(x\right)|\nabla\psi|dV_{S}.
\end{equation}
This is the weighted counterpart of the convex relaxation in ref.
\cite{Headrick:2017ucz}.

To identify this convex program with a weighted minimal cut, define
the one-parameter family of bulk regions
\begin{equation}
r\left(p\right):=\left\{ x\in\Sigma:\psi\left(x\right)\geq p\right\} ,\qquad p\in\mathbb{R},
\end{equation}
 and let bulk surface
\begin{equation}
m\left(p\right):=\overline{\partial r\left(p\right)\setminus\partial\Sigma},
\end{equation}
 be the corresponding bulk level set. The weighted coarea formula
gives 
\begin{equation}
\int_{\Sigma}\rho_{S}\left(x\right)|\nabla\psi|dV_{S}=\int_{-\infty}^{\infty}dp\int_{m\left(p\right)}\rho_{S}\left(x\right)dA_{S}.
\end{equation}
For $0<p<1$, the boundary conditions on $\psi$ imply $r\left(p\right)\cap\left(\partial\Sigma\setminus R\right)=A$
and thus $m\left(p\right)\sim A$ rel $R$. Since $\rho_{S}\left(x\right)$
is positive, it follows that

\begin{equation}
\int_{\Sigma}\rho_{S}|\nabla\psi|dV_{S}\geq\int_{0}^{1}dp\int_{m\left(p\right)}\rho_{S}dA_{S}\geq\inf_{\substack{m\sim A\\
\mathrm{rel}\ R
}
}\int_{m}\rho_{S}dA_{S}.
\end{equation}
Taking the infimum over $\psi$ therefore gives 
\begin{equation}
\inf_{\substack{\psi|_{A}=1,\ \psi|_{\bar{A}}=0}
}\int_{\Sigma}\rho_{S}|\nabla\psi|dV_{S}\geq\inf_{\substack{m\sim A\\
\mathrm{rel}\ R
}
}\int_{m}\rho_{S}dA_{S}.
\end{equation}
Conversely, consider any cut $m\sim A$, and let $r$ be a bulk region
whose boundary is $A-m$, up to a possible component on $R$. The
characteristic function $\chi_{r}$ can be approximated by regular
scalar fields whose transition layers become arbitrarily thin around
$m$. In this limit,

\begin{equation}
\int_{\Sigma}\rho_{S}|\nabla\psi|dV_{S}\longrightarrow\int_{m}\rho_{S}dA_{S}.
\end{equation}
It follows that

\begin{equation}
\inf_{\substack{\psi|_{A}=1,\ \psi|_{\bar{A}}=0}
}\int_{\Sigma}\rho_{S}|\nabla\psi|dV_{S}=\inf_{\substack{m\sim A\\
\mathrm{rel}\ R
}
}\int_{m}\rho_{S}dA_{S}.
\end{equation}
Combining this result with strong convex duality yields the weighted
Riemannian max-flow/min-cut relation,

\begin{equation}
\sup_{\substack{\nabla_{i}v^{i}=0\\
|v|\leq\rho_{S}\\
v^{i}n_{i}|_{R}=0
}
}\int_{A}v^{i}n_{i}dA_{S}=\inf_{\substack{m\sim A\\
\mathrm{rel}\ R
}
}\int_{m}\rho_{S}dA_{S}.
\end{equation}
When the optimal flow and cut exist, the supremum and infimum may
be replaced by a maximum and a minimum. Substituting the sphere-derived
string-frame capacity, $\rho_{S}\left(x\right)$, we obtain

\begin{equation}
\min_{\substack{m\sim A\\
\mathrm{rel}\ R
}
}\frac{1}{4G_{N}}\int_{m}e^{-2\phi\left(x\right)}dA_{S}=\max_{\substack{\nabla_{i}v^{i}=0\\
|v|\leq\rho_{S}\\
v^{i}n_{i}|_{R}=0
}
}\int_{A}v^{i}n_{i}dA_{S}.
\end{equation}
Together with the weighted minimal-cut formula derived from the genus-zero
off-shell string functional, this gives

\begin{equation}
S_{A}^{\left(0\right)}=\max_{v}\Phi\left[A\right]=\max_{v}\int_{A}v^{i}n_{i}dA_{S},
\end{equation}
subject to
\begin{equation}
\nabla_{i}v^{i}=0,\qquad\left|v\right|\leq\frac{e^{-2\phi\left(x\right)}}{4G_{N}},\qquad v^{i}n_{i}|_{R}=0\quad\text{if }R\neq\varnothing.
\end{equation}
Thus the weighted flow--cut duality yields the string-frame norm
bound from the replica-selected entropy functional. Moreover, for
a prescribed stationary horizon $\Gamma_{H}$ which is the relevant
bottleneck, the maximal entropy flux becomes

\begin{align}
\Phi\left(\Gamma_{H}\right) & =\int_{\Gamma_{H}}v^{i}n_{i}dA_{S}\nonumber \\
 & =\int_{\Gamma_{H}}\rho_{S}dA_{S}=\frac{\mathcal{A}_{\phi}\left(\Gamma_{H}\right)}{4G_{N}}=S_{{\rm BH}}^{\left(0\right)}.
\end{align}
This is the precise relation between the Ahmadain--Wall generalized
area and the flow bound: $\mathcal{A}_{\phi}/\left(4G_{N}\right)$
is the integrated capacity of the horizon, and $\rho_{S}\left(x\right)=e^{-2\phi}/\left(4G_{N}\right)$
is the capacity per unit string-frame area. In the BTZ application,
the horizon is a flux-carrying terminal or a homologous interior cut,
and is not included in the impermeable component $R$.

\subsection{Physical interpretations}

We now summarize the physical interpretations of the results presented
above. At genus zero and to two-derivative order, the regulated sphere
functional determines the tree-level target-space action through

\begin{equation}
Z_{0}=\frac{\partial K_{0}}{\partial\log\epsilon},\qquad I_{0}=-Z_{0}.
\end{equation}
The conical variation of this action gives the local string-frame
entropy functional

\begin{equation}
S_{{\rm loc}}^{\left(0\right)}\left[m\right]=\int_{m}\rho_{S}\left(x\right)dA_{S},\qquad\rho_{S}\left(x\right)=\frac{e^{-2\phi\left(x\right)}}{4G_{N}}.
\end{equation}
The gravitational replica construction then selects the weighted minimal
cut,

\begin{equation}
S_{A}^{\left(0\right)}=\min_{m\sim A}\int_{m}\rho_{S}\left(x\right)dA_{S},
\end{equation}
and the weighted Riemannian max-flow/min-cut theorem gives

\begin{equation}
S_{A}^{\left(0\right)}=\max_{v}\int_{A}v^{i}n_{i}dA_{S},\qquad\nabla_{i}v^{i}=0,\qquad\left|v\right|\leq\frac{e^{-2\phi\left(x\right)}}{4G_{N}}.
\end{equation}
In this qualified but precise sense, the string-frame norm bound is
derived from the genus zero string sphere through the following sequence:

\begin{equation}
\begin{array}{c}
S_{{\rm loc}}^{\left(0\right)}\left[m\right]\\
\mathrm{off-shell\;string}
\end{array}\longrightarrow\begin{array}{c}
S_{A}^{\left(0\right)}\\
\mathrm{gravi-replica}
\end{array}\longrightarrow\begin{array}{c}
\left|v\right|\leq\frac{e^{-2\phi\left(x\right)}}{4G_{N}}\\
\mathrm{max-flow/min-cut}
\end{array}
\end{equation}
In other words, the off-shell worldsheet calculation first fixes the
local entropy density $\rho_{S}\left(x\right)$. The gravitational
replica construction then converts this density into a weighted minimal
cut functional. Finally, convex duality converts the resulting surface
functional into the local capacity constraint

\begin{equation}
\left|v\right|\leq\frac{e^{-2\phi\left(x\right)}}{4G_{N}}.
\end{equation}
This is the principal physical content of the construction. Flow conservation,

\begin{equation}
\nabla_{i}v^{i}=0,
\end{equation}
is the dual constraint associated with the unrestricted interior values
of the scalar variable $\psi$. It guarantees conservation of the
flow but does not determine its magnitude. The norm bound gives the
local gravitational information: it contains the Newton constant,
the dilaton function, and the maximal entropy flux that can pass through
a unit string-frame area. Thus,

\begin{equation}
\rho_{S}\left(x\right)=\frac{e^{-2\phi\left(x\right)}}{4G_{N}}
\end{equation}
has the direct interpretation of an entropy capacity.

This interpretation also gives a precise flow meaning to the Ahmadain--Wall
generalized area,

\begin{equation}
\mathcal{A}_{\phi}\left(m\right):=\int_{m}e^{-2\phi\left(x\right)}dA_{S}.
\end{equation}
Its relation to the flow capacity is

\begin{equation}
\frac{\mathcal{A}_{\phi}\left(m\right)}{4G_{N}}=\int_{m}\rho_{S}\left(x\right)dA_{S}.
\end{equation}
Consequently, $\mathcal{A}_{\phi}/\left(4G_{N}\right)$ is the integrated
entropy capacity of a cut, whereas $\rho_{S}\left(x\right)$ is the
corresponding capacity per unit string-frame area. For a stationary
black hole whose bifurcation surface $\Gamma_{H}$ is the relevant
bottleneck,

\begin{equation}
\max_{v}\Phi\left[\Gamma_{H}\right]=\frac{1}{4G_{N}}\int_{\Gamma_{H}}e^{-2\phi\left(x\right)}\mathrm{d}A_{S}=\frac{\mathcal{A}_{\phi}\left(\Gamma_{H}\right)}{4G_{N}}=S_{\mathrm{BH}}^{\left(0\right)}.
\end{equation}
The same construction applies to a general static boundary region
$A$. \emph{The black hole entropy and the entanglement entropy are
therefore governed by the same capacity density, while the relevant
homology condition determines the corresponding bottleneck.}

The dilaton factor is also precisely the factor required by the relation
between the string and Einstein frames. In $D$ spacetime dimensions,

\begin{equation}
g_{\mu\nu}^{E}=e^{-\frac{4\phi}{D-2}}g_{\mu\nu}^{S},
\end{equation}
and the induced codimension-two area elements obey

\begin{equation}
dA_{E}=e^{-2\phi}dA_{S}.
\end{equation}
It follows immediately that

\begin{equation}
\frac{dA_{E}}{4G_{N}}=\frac{e^{-2\phi}dA_{S}}{4G_{N}}.
\end{equation}
More explicitly, introducing

\begin{equation}
\Omega=e^{-\frac{2\phi}{D-2}},\qquad h_{ij}^{E}=\Omega^{2}h_{ij}^{S},
\end{equation}
the flux-preserving relation between the two flow fields is
\begin{equation}
v_{S}^{i}=\Omega^{D-1}v_{E}^{i}.
\end{equation}
Indeed,

\begin{equation}
\sqrt{h_{S}}v_{S}^{i}=\sqrt{h_{E}}v_{E}^{i},
\end{equation}
so the integrated flux is frame invariant. Moreover,
\begin{equation}
\nabla_{i}^{S}v_{S}^{i}=0\quad\Longleftrightarrow\quad\nabla_{i}^{E}v_{E}^{i}=0,
\end{equation}
and their norms are related by

\begin{equation}
|v_{S}|=\Omega^{D-2}|v_{E}|=e^{-2\phi}|v_{E}|.
\end{equation}
Hence the standard Einstein-frame norm bound

\begin{equation}
|v_{E}|\leq\frac{1}{4G_{N}},
\end{equation}
is equivalent to the string-frame bound 
\begin{equation}
|v_{S}|\leq\frac{e^{-2\phi}}{4G_{N}}.
\end{equation}
The sphere derivation explains why this frame transformation appears
locally in the flow description: the factor $e^{-2\phi}$ originates
from the genus-zero closed-string measure and therefore accompanies
the entropy capacity in string-frame variables.

We now relate this geometrically determined flow capacity to the projected
string charge density. Varying the worldsheet coupling to the Kalb--Ramond
field produces the antisymmetric spacetime current $j^{\mathrm{ab}}$.
On a stationary constant-time slice, its projected spatial string
charge density is

\begin{equation}
q^{i}=N_{\mathrm{ADM}}j^{0i}=\frac{\sqrt{-g}}{\sqrt{h}}j^{0i},
\end{equation}
and current conservation implies
\begin{equation}
\partial_{i}\left(\sqrt{h}q^{i}\right)=0.
\end{equation}
The projected string charge density therefore has precisely the conservation
property required to define candidate flow lines. The proposed string
charge density/bit threads correspondence introduces the additional
identification
\begin{equation}
v^{i}=C_{\mathrm{geom}}q^{i},
\end{equation}
Under this identification, the sphere-derived geometric capacity imposes
the constraint

\begin{equation}
C_{\mathrm{geom}}\left|q\left(x\right)\right|\leq\rho_{S}\left(x\right)=\frac{e^{-2\phi\left(x\right)}}{4G_{N}}.
\end{equation}
For the homology class under consideration, conservation makes the
string-charge flux cut independent:
\begin{equation}
\mathcal{N}=\int_{m}q^{i}n_{i}dA_{S},\qquad m\sim A.
\end{equation}
The capacity bound then implies
\begin{equation}
C_{\mathrm{geom}}\mathcal{N}=C_{\mathrm{geom}}\int_{m}q^{i}n_{i}dA_{S}\leq\int_{m}\rho_{S}\left(x\right)\mathrm{d}A_{S}.
\end{equation}
Since this inequality holds for every cut $m\sim A$, minimization
gives
\begin{equation}
C_{\mathrm{geom}}\mathcal{N}\leq\min_{m\sim A}\int_{m}\rho_{S}\left(x\right)\mathrm{d}A_{S}=S_{A}^{\left(0\right)}.
\end{equation}
When the projected string current realizes the corresponding maximal
flow,
\begin{equation}
C_{\mathrm{geom}}\mathcal{N}=S_{A}^{\left(0\right)},
\end{equation}
At the minimizing cut $m_{A}$, the two inequalities above are simultaneously
saturated. Consequently, the flow is normal to the cut and saturates
the local capacity:
\begin{equation}
\left.v^{i}\right|_{m_{A}}=\left.\rho_{S}n^{i}\right|_{m_{A}},
\end{equation}
Using $v^{i}=C_{\mathrm{geom}}q^{i}$, this gives

\begin{equation}
\left.C_{\mathrm{geom}}\left|q\left(x\right)\right|\right|_{m_{A}}=\left.\rho_{S}\left(x\right)\right|_{m_{A}}=\left.\frac{e^{-2\phi\left(x\right)}}{4G_{N}}\right|_{m_{A}}.
\end{equation}
For a black hole bifurcation surface, this becomes

\begin{equation}
C_{\mathrm{geom}}\mathcal{N}=S_{\mathrm{BH}}^{\left(0\right)},
\end{equation}
when the projected physical string charge density provides the maximal-flow
representative.

\section{Physical string sources and dynamical selection of a maximal flow}

\label{sec:physical-string-BTZ}

The results in the previous sections determine the string-frame entropy
capacity, or norm bound 
\begin{equation}
\rho_{S}\left(x\right)=\frac{e^{-2\phi\left(x\right)}}{4G_{N}},\qquad\nabla_{i}v^{i}=0,\qquad|v|\leq\rho_{S}\left(x\right).\label{eq:phys-capacity}
\end{equation}
We now construct a dynamical F1 realization of the corresponding maximal
flow. In the classical probe regime adopted here, the BTZ fields are
held fixed while the string embedding is determined by the full classical
F1 equations, 
\begin{equation}
\frac{\delta S_{\mathrm{F1}}}{\delta X^{\mathrm{a}}}=0.
\end{equation}
In this precise sense, a physical string denotes an on-shell F1 probe.
Varying its Wess-Zumino coupling with respect to $B_{ab}$ produces
a genuine Kalb-Ramond charge current and density, so the worldsheet
dynamics determines both the physical current and its spatial direction.
For the nonrotating BTZ background supported by NS-NS flux, we show
that the regular stationary embedding produces a conserved current
whose capacity-weighted normalization realizes a maximal entropy flow.
A uniform distribution of such strings provides the smooth coarse-grained
flow in the following calculations. It is worth noting that the uniform
smearing is an ensemble assumption, while the individual string shapes
follow from their equations of motion. The continuum approximation
and the probe approximation are separate assumptions. Specifically,
we neglect both the background response to the ensemble and interactions
among its members. We use a regime in which these approximations are
compatible, rather than sending $\mathcal{N}$ to infinity at fixed
gravitational coupling.

\subsection{Stationary strings in the NS--NS BTZ background}

We consider the three-dimensional NS--NS action 
\begin{equation}
S_{\mathrm{bulk}}^{\left(0\right)}=\frac{1}{16\pi G_{N}^{\left(3\right)}}\int d^{3}x\sqrt{-g}e^{-2\phi}\left[R+4\left(\nabla\phi\right)^{2}-\frac{1}{12}H^{2}+\frac{4}{\ell^{2}}\right],\qquad H=dB,
\end{equation}
with $H=\left(2/\ell\right)\varepsilon_{\mathrm{BTZ}}$. The BTZ black
hole solution is \cite{Banados:1992wn,Horowitz:1993jc}
\begin{align}
dS_{\mathrm{BTZ}}^{2} & =-f\left(r\right)dt^{2}+\frac{dr^{2}}{f\left(r\right)}+r^{2}d\theta^{2}, & f\left(r\right) & =\frac{r^{2}-r_{h}^{2}}{\ell^{2}}, & \theta & \sim\theta+2\pi,\\
\phi & =\phi_{0}, & B & =-b\left(r\right)dt\wedge d\theta, & b\left(r\right) & =\frac{r^{2}-r_{h}^{2}}{\ell}.\label{eq:phys-BTZ-background}
\end{align}
The additive constant in $b$ is chosen so that the potential is regular
at the horizon. The orientation of the NS--NS flux is fixed by this
convention. 

\begin{figure}[h]
\begin{centering}
\includegraphics[scale=0.4]{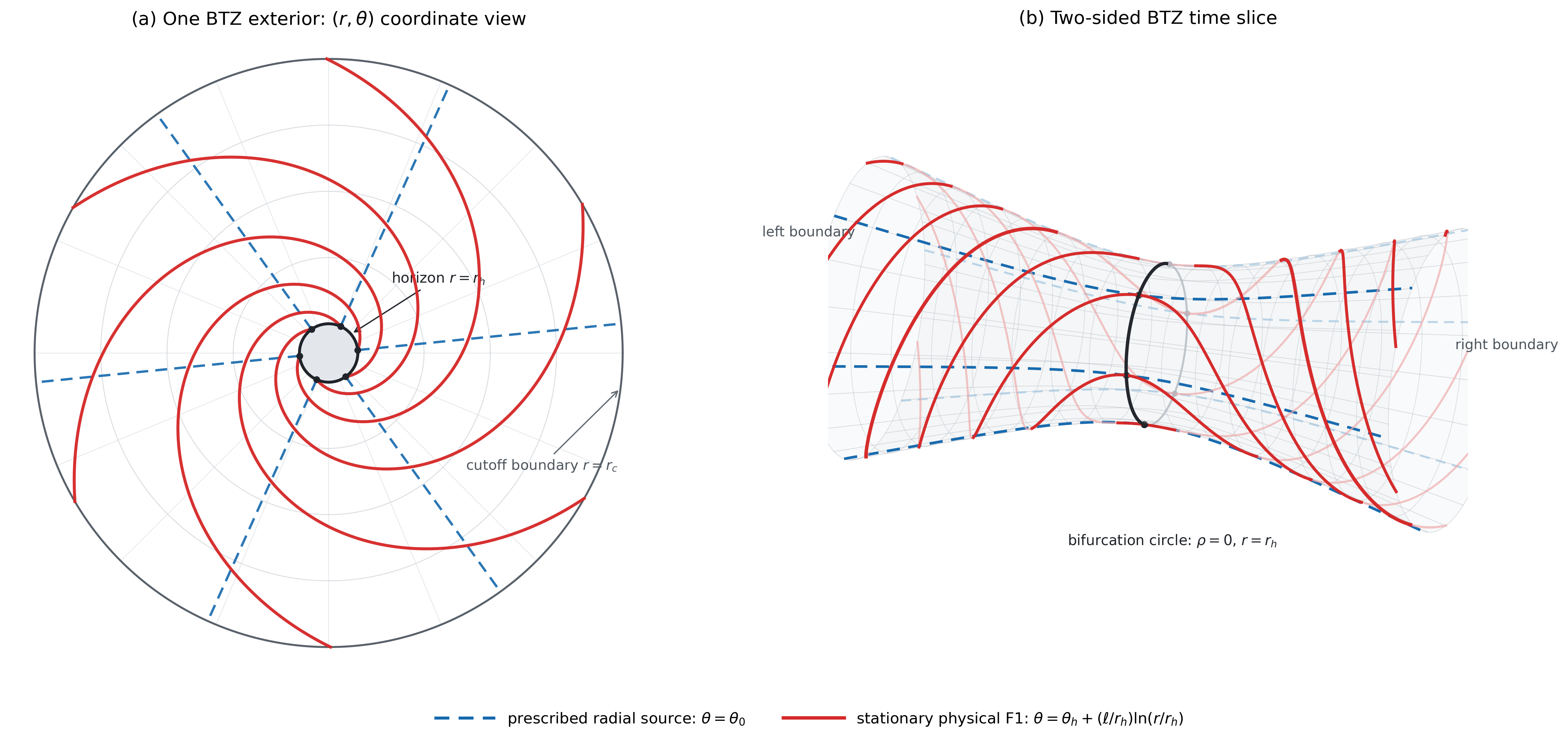}
\par\end{centering}
\caption{\label{fig:new source}Comparison between the prescribed radial string
sources and the stationary physical F1 sources on a constant-time
slice of one BTZ exterior and two-sided BTZ geometry. The blue dashed
curves represent the radially embedded sources $X^{\mu}=\left(\tau,r,\theta_{0}\right)$,
whereas the red solid curves represent the stationary spiral solutions
$X^{\mu}=\left(\tau,r,\theta\left(r\right)\right)$. The latter acquire
an angular component due to the background NS--NS flux but become
purely normal at the horizon, where $v_{\mathrm{phys}}^{i}=\rho_{S}n^{i}$. }
\end{figure}

The classical F1 action is given by \cite{Dabholkar:1990yf,Sen:1992yt,Duff:1994an}

\begin{equation}
S_{\mathrm{F1}}=-\frac{1}{4\pi\alpha^{\prime}}\int d^{2}\sigma\left(\sqrt{-\gamma}\gamma^{mn}\partial_{m}X^{\mathrm{a}}\partial_{n}X^{\mathrm{b}}g_{\mathrm{ab}}+\epsilon^{mn}\partial_{m}X^{\mathrm{a}}\partial_{n}X^{\mathrm{b}}B_{\mathrm{ab}}\right),
\end{equation}
where $\gamma_{mn}$ is the auxiliary worldsheet metric. Eliminating
it gives the Nambu--Goto form with induced metric $\gamma_{mn}^{\mathrm{ind}}=g_{\mathrm{ab}}\partial_{m}X^{\mathrm{a}}\partial_{n}X^{\mathrm{b}}$.
The constant dilaton gives no local contribution to the classical
embedding equations. For a stationary exterior string, we choose static
gauge 

\begin{equation}
X^{t}=\tau,\qquad X^{r}=\sigma,\qquad X^{\theta}=\theta\left(r\right).
\end{equation}
Using the notations $\epsilon^{\tau\sigma}=+1$, $\theta^{\prime}\left(r\right)\equiv d\theta/dr$,
yields 

\begin{equation}
-\det\gamma_{mn}^{\mathrm{ind}}=1+fr^{2}\theta^{\prime2},\qquad S_{\mathrm{F1}}=-\frac{1}{2\pi\alpha^{\prime}}\int dtdr\left(\sqrt{1+fr^{2}\theta^{\prime2}}-b\theta^{\prime}\right).
\end{equation}
Since $\theta$ is cyclic, its equation of motion has the first integral
\begin{equation}
\frac{f\left(r\right)r^{2}\theta^{\prime}\left(r\right)}{\sqrt{1+f\left(r\right)r^{2}\theta^{\prime2}\left(r\right)}}-b\left(r\right)=k_{\theta}.\label{eq:angular integral}
\end{equation}
The remaining embedding equations follow from worldsheet reparametrization
invariance once the transverse equation and the auxiliary-metric equation
hold. A local constant gauge shift $b\to b+b_{0}$ shifts $k_{\theta}\to k_{\theta}-b_{0}$,
leaving the combination $b+k_{\theta}$ and the embedding unchanged.
Horizon regularity fixes $k_{\theta}=-b\left(r_{h}\right)$, which
is zero in the regular potential gauge. We select the branch on which
$\theta\left(r\right)$ is smooth with finite $\theta^{\prime}$ at
$r=r_{h}$. Since $f\left(r_{h}\right)=b\left(r_{h}\right)=0$, this
branch has $k_{\theta}=0$. This is a specified regularity condition.
Solving Eq. (\ref{eq:angular integral}) gives 
\begin{equation}
\theta^{\prime2}\left(r\right)=\frac{b^{2}}{fr^{2}\left(fr^{2}-b^{2}\right)},\qquad fr^{2}-b^{2}=\frac{r_{h}^{2}\left(r^{2}-r_{h}^{2}\right)}{\ell^{2}},
\end{equation}
and hence, for our orientation, 

\begin{equation}
\theta^{\prime}\left(r\right)=\frac{\ell}{r_{h}r},\qquad\theta\left(r\right)=\theta_{h}+\frac{\ell}{r_{h}}\log\frac{r}{r_{h}},\qquad\sqrt{-\det\gamma^{\mathrm{ind}}}=\frac{r}{r_{h}}.\label{eq:phys spiral solution}
\end{equation}
The solution is regular at the horizon. Its angular winding grows
without bound as $r\to\infty$, so an individual curve has no fixed
angular endpoint at the conformal boundary. We define the ensemble
and its endpoint data at a finite radial cutoff $r_{c}$. Thus, the
conserved entropy flux found below is independent of $r_{c}$. This
result gives a regular on-shell F1 embedding in the exterior. Both
$\theta^{\prime}\left(r\right)$ and the induced worldsheet determinant
remain finite at the horizon, and the solution satisfies the complete
classical embedding equations. The logarithmic spiral is thus dynamically
selected by the balance between the string tension and the coupling
to the background NS-NS flux. Its projected Kalb-Ramond charge density
then provides the physical current. Moreover, the difference between
prescribed radial string sources and the stationary on-shell F1 sources
on a constant time slice of one BTZ exterior and two-sided BTZ geometry
is illustrated in Fig. (\ref{fig:new source})

Classical string dynamics in BTZ backgrounds, including fundamental
string probes in the presence of NS--NS flux, has been studied previously
\cite{Larsen:1994ah,Ashok:2021ffx}. Our purpose here is to relate
the regular stationary branch to its projected charge density and
to the independently determined entropy capacity.

The distinction from the purely radial source is essential. Setting
$\theta^{\prime}=0$ in Eq. (\ref{eq:angular integral}) would require
$b^{\prime}=0$, whereas here 

\begin{equation}
b^{\prime}\left(r\right)=\frac{2r}{\ell}\neq0.
\end{equation}
Although the term $\epsilon^{mn}\partial_{m}X^{\mathrm{a}}\partial_{n}X^{\mathrm{b}}B_{\mathrm{ab}}$
in action vanishes on a radial embedding, its first variation does
not: 
\begin{equation}
\left.\delta S_{B}\right|_{\theta^{\prime}=0}=-\frac{1}{2\pi\alpha^{\prime}}\int dtdrb^{\prime}\left(r\right)\delta\theta+\text{boundary terms}.
\end{equation}
Consequently, the radially embedded current used in our previous construction, 

\begin{equation}
X^{\mu}=\left(\tau,r,\theta_{0}\right),
\end{equation}
defines a conserved radial charge density, but it is not a free stationary
F1 solution of the embedding equations in the present NS-NS background.
It may instead be interpreted as a fixed external source, or instantaneous
initial data for a time-dependent string. This distinction concerns
the source dynamics: the prescribed radial current remains conserved
and, with the capacity normalization, still defines a maximal entropy
flow. The stationary strings constructed here provide an on-shell
representative with the same maximal entropy flux.

\subsection{The physical string current and black hole entropy}

We normalize the antisymmetric charge current so that a single oriented
string carries one unit of charge. With $\epsilon^{\tau\sigma}=+1$,
its distributional expression is 
\begin{equation}
j^{\mathrm{ab}}\left(x\right)=\frac{1}{\sqrt{-g\left(x\right)}}\sum_{k=1}^{\mathcal{N}}\int d\tau d\sigma\epsilon^{mn}\partial_{m}X_{k}^{a}\partial_{n}X_{k}^{b}\delta^{\left(3\right)}\left(x-X_{k}\right).\label{eq:phys current}
\end{equation}
The physical source appearing in the variation of the $B$-coupling
contains the additional factor $1/\left(2\pi\alpha^{\prime}\right)$,
and this factor has been divided out in Eq. (\ref{eq:phys current}).
Thus $\mathcal{N}$ denotes normalized string charge and counts strings
only when all sources have the same orientation and unit charge. 

Consider the ensemble of Eq. (\ref{eq:phys spiral solution}), with
$\theta_{h}$ distributed uniformly around the circle. In the continuum
approximation, 
\begin{equation}
\sum_{k}\delta_{2\pi}\left(\theta-\theta_{k}\left(r\right)\right)\longrightarrow\frac{\mathcal{N}}{2\pi}.
\end{equation}
Since $\sqrt{-g}=r$, its nonzero independent current components are
\begin{equation}
j^{0r}=\frac{\mathcal{N}}{2\pi r},\qquad j^{0\theta}=\frac{\mathcal{N}}{2\pi r}\theta^{\prime}\left(r\right),\qquad j^{r\theta}=0.\label{eq:phys spacetime current}
\end{equation}
The induced metric and lapse on a static exterior slice $\Sigma$
are 
\begin{equation}
h_{ij}dx^{i}dx^{j}=\frac{dr^{2}}{f}+r^{2}d\theta^{2},\qquad\sqrt{h}=\frac{r}{\sqrt{f}},\qquad N_{\mathrm{ADM}}=\sqrt{f}.
\end{equation}
Projection with the future-directed unit normal $n_{\mu}$ gives $q^{\mu}=-n_{\nu}j^{\nu\mu}$,
and therefore 
\begin{eqnarray}
q^{r} & = & N_{\mathrm{ADM}}j^{0r}=\frac{\mathcal{N}\sqrt{f}}{2\pi r},\nonumber \\
q^{\theta} & = & N_{\mathrm{ADM}}j^{0\theta}=\frac{\mathcal{N}\ell\sqrt{f}}{2\pi r_{h}r^{2}}=\theta^{\prime}q^{r}.
\end{eqnarray}
The slice projection retains the angular component. Current conservation
and the flux through a constant-$r$ circle are verified directly: 

\begin{eqnarray}
\nabla_{i}q^{i} & = & \frac{1}{\sqrt{h}}\left[\partial_{r}\left(\sqrt{h}q^{r}\right)+\partial_{\theta}\left(\sqrt{h}q^{\theta}\right)\right]\nonumber \\
 & = & \frac{\sqrt{f}}{r}\partial_{r}\left[\frac{r}{\sqrt{f}}\frac{\mathcal{N}\sqrt{f}}{2\pi r}\right]\nonumber \\
 & = & \frac{\sqrt{f}}{r}\partial_{r}\left(\frac{\mathcal{N}}{2\pi}\right)=0,
\end{eqnarray}
and 
\begin{equation}
\int_{r=\mathrm{const}}q^{i}n_{i}dA_{S}=\mathcal{N}.
\end{equation}
Here the circle's outward unit normal is $n^{i}=\sqrt{f}\left(\partial_{r}\right)^{i}$
and $dA_{S}=rd\theta$. The angular derivative vanishes because the
ensemble is uniformly distributed. More significantly, the norm of
the full current is constant: 
\begin{align}
|q|^{2} & =h_{ij}q^{i}q^{j}=\frac{\left(q^{r}\right)^{2}}{f}+r^{2}\left(q^{\theta}\right)^{2}=\left(\frac{\mathcal{N}}{2\pi r}\right)^{2}\left(1+fr^{2}\theta^{\prime2}\right)=\left(\frac{\mathcal{N}}{2\pi r_{h}}\right)^{2}.\label{eq:phys current norm}
\end{align}
Therefore, we successfully obtain the constant norm that comes from
the stationary embedding equation and the uniform distribution, rather
than an imposed choice of radial flow profile.

We now define the unit-normalized current

\begin{equation}
\widehat{q}^{i}\equiv\frac{q^{i}}{|q|}.
\end{equation}
Its components are 
\begin{equation}
\widehat{q}^{r}=\frac{r_{h}\sqrt{f}}{r},\qquad\widehat{q}^{\theta}=\frac{\ell\sqrt{f}}{r^{2}},
\end{equation}
 and it satisfies 
\begin{equation}
|\widehat{q}|=1.
\end{equation}
The constancy of $|q|$ is crucial. For a general conserved vector
field,

\begin{equation}
\nabla_{i}q^{i}=0,
\end{equation}
normalization does not necessarily preserve conservation. In fact,

\begin{equation}
\nabla_{i}\left(\frac{q^{i}}{|q|}\right)=\frac{\nabla_{i}q^{i}}{|q|}-\frac{q^{i}\partial_{i}|q|}{|q|^{2}}.
\end{equation}
Consequently, when $\nabla_{i}q^{i}=0$, 
\begin{equation}
\nabla_{i}\left(\frac{q^{i}}{|q|}\right)=-\frac{q^{i}\partial_{i}|q|}{|q|^{2}},
\end{equation}
which is generally nonzero. For the stationary spiral solution, however,
\begin{equation}
\partial_{i}|q|=0.
\end{equation}
Therefore, 
\begin{equation}
\nabla_{i}\widehat{q}^{i}=\nabla_{i}\left(\frac{q^{i}}{|q|}\right)=0.
\end{equation}
It implies that the physical F1 equation gives more than a conserved
charge current: it produces a current whose unit normalization remains
divergenceless.

We next compute the flux of the unit current. For a constant-$r$
circle $C_{r}$, the outward unit normal and induced line element
are

\begin{equation}
n^{i}=\sqrt{f}\left(\partial_{r}\right)^{i},\qquad n_{i}dx^{i}=\frac{dr}{\sqrt{f}},\qquad dA_{S}=rd\theta.
\end{equation}
It follows that 
\begin{equation}
\widehat{q}^{i}n_{i}=\frac{\widehat{q}^{r}}{\sqrt{f}}=\frac{r_{h}}{r}.
\end{equation}
Therefore,

\begin{equation}
\Phi\left[\widehat{q}\right]=\int_{0}^{2\pi}\widehat{q}^{i}n_{i}dA_{S}=2\pi r_{h}=A_{H},
\end{equation}
which implies the flux of the unit-normalized physical string current
equals the string-frame circumference of the BTZ horizon. %
We now recall the string-frame norm bound derived from the off-shell
sphere and replica construction:

\begin{equation}
\rho_{S}\left(x\right)=\frac{e^{-2\phi_{0}}}{4G_{N}^{\left(3\right)}}.
\end{equation}
Because the dilaton is constant in the present BTZ background, $\rho_{S}$
is also constant. Combining this independently derived norm bound
with the dynamically determined unit direction gives the entropy flow

\begin{equation}
v_{\mathrm{phys}}^{i}\equiv\rho_{S}\widehat{q}^{i}=\rho_{S}\frac{q^{i}}{|q|}.\label{eq:phys capacity}
\end{equation}
This is the important relation that shows the correspondence between
the physical string charge density and bit threads $v_{\mathrm{phys}}^{i}$,
where $\rho_{S}$ derived from off-shell string theory. Its explicit
form is

\begin{equation}
v_{\mathrm{phys}}=\rho_{S}\frac{r_{h}\sqrt{f}}{r}\partial_{r}+\rho_{S}\frac{\ell\sqrt{f}}{r^{2}}\partial_{\theta}.
\end{equation}
Since $\rho_{S}$ is constant and $\widehat{q}^{i}$ is divergenceless,
we can obtain the divergenceless condition for $v_{\mathrm{phys}}$:

\begin{equation}
\nabla_{i}v_{\mathrm{phys}}^{i}=\rho_{S}\nabla_{i}\widehat{q}^{i}=0.
\end{equation}
Furthermore, 
\begin{equation}
|v_{\mathrm{phys}}|=\rho_{S}|\widehat{q}|=\rho_{S}.
\end{equation}
Thus, we have
\begin{equation}
|v_{\mathrm{phys}}|=\frac{e^{-2\phi_{0}}}{4G_{N}^{\left(3\right)}},
\end{equation}
throughout the exterior region. The physical string flow saturates
the string-frame norm bound everywhere. Its flux is 

\begin{equation}
\Phi\left[v_{\mathrm{phys}}\right]=\rho_{S}\Phi\left[\widehat{q}\right]=\frac{e^{-2\phi_{0}}\left(2\pi r_{h}\right)}{4G_{N}^{\left(3\right)}}=S_{\mathrm{BH}}^{\left(0\right)}.
\end{equation}
In other words

\begin{equation}
S_{\mathrm{BH}}^{(0)}=\int v_{\mathrm{phys}}^{i}n_{i}dA_{S}=\frac{e^{-2\phi_{0}}}{4G_{N}^{\left(3\right)}}\int\frac{q^{i}}{|q|}n_{i}dA_{S}.
\end{equation}

On the other hand, it remains to verify that $v_{\mathrm{phys}}^{i}$
is a maximal flow. At the horizon, its orthonormal components are

\begin{equation}
v_{\mathrm{phys}}^{\hat{r}}\left(r\right)=\frac{v_{\mathrm{phys}}^{r}}{\sqrt{f}}=\rho_{S}\frac{r_{h}}{r},\qquad v_{\mathrm{phys}}^{\hat{\theta}}\left(r\right)=rv_{\mathrm{phys}}^{\theta}=\rho_{S}\frac{\sqrt{r^{2}-r_{h}^{2}}}{r}.
\end{equation}
Therefore, we have

\begin{equation}
v_{\mathrm{phys}}^{i}\left(r_{h}\right)=\rho_{S}n^{i},
\end{equation}
on the horizon. It implies

\begin{equation}
\Phi\left[v_{\mathrm{phys}}\right]=\rho_{S}\Phi\left[\widehat{q}\right]=2\pi r_{h}\rho_{S},\qquad|v_{\mathrm{phys}}^{i}\left(r_{h}\right)|=\rho_{S}
\end{equation}
The flow is normal to the bottleneck and saturates the capacity there.
For any admissible flow $w^{i}$ in the same homology class,

\begin{equation}
\nabla_{i}w^{i}=0,\qquad|w|\leq\rho_{S}.
\end{equation}
Its flux is bounded by the horizon cut:

\begin{equation}
\Phi\left[w\right]=\int_{\Gamma_{H}}w\cdot ndA_{S}\leq\int_{\Gamma_{H}}|w|dA_{S}\leq\int_{\Gamma_{H}}\rho_{S}dA_{S}=2\pi r_{h}\rho_{S}.
\end{equation}
Therefore, $v_{\mathrm{phys}}^{i}$ is a global maximal flow:

\begin{equation}
\Phi\left[w\right]\leq2\pi r_{h}\rho_{S}=\Phi\left[v_{\mathrm{phys}}\right],\qquad|w|\leq|v_{\mathrm{phys}}^{i}\left(r_{h}\right)|.
\end{equation}
The horizon is also a weighted minimal cut. For any homologous cut
$m$, conservation and admissibility give 
\begin{equation}
\int_{m}\rho_{S}dA_{S}\geq\int_{m}v_{\mathrm{phys}}^{i}n_{i}dA_{S}=2\pi r_{h}\rho_{S}=\int_{\Gamma_{H}}\rho_{S}dA_{S}.
\end{equation}
This argument does not restrict competing cuts to circles. The entropy
problem here concerns the complete boundary of a BTZ exterior, with
the horizon as the other terminal. Its flux is the thermal entropy
of the BTZ saddle, or equivalently the entropy of either boundary
CFT in the pure thermofield-double description of two-sided BTZ \cite{Maldacena:2001kr}.
Note that a radial cutoff regulates the string endpoints, and the
smooth horizon is a crossing surface, not a physical endpoint of the
string.

\subsection{Comparison with the original string charge density/bit threads correspondence}

We can now compare the present construction with the original string
charge density/bit threads correspondence,

\begin{equation}
v^{i}=C_{\mathrm{geom}}q^{i},\qquad C_{\mathrm{geom}}>0.
\end{equation}
where $q^{i}$ denotes the unnormalized string-charge current. For
the stationary BTZ string ensemble constructed above, its norm is
spatially constant:

\begin{equation}
|q|=\frac{\mathcal{N}}{2\pi r_{h}}.
\end{equation}
The corresponding unit-normalized current is 
\begin{equation}
\widehat{q}^{i}=\frac{q^{i}}{|q|}.
\end{equation}
Identifying the flow $v^{i}$ in the original correspondence with
the particular maximal-flow representative constructed from the stationary
physical-string ensemble, 
\begin{equation}
v^{i}\equiv v_{\mathrm{phys}}^{i},
\end{equation}
we have

\begin{equation}
v_{\mathrm{phys}}^{i}=\rho_{S}\hat{q}^{i}=\rho_{S}\frac{q^{i}}{|q|}.
\end{equation}
The same flow also can be written as

\begin{equation}
v_{\mathrm{phys}}^{i}=\left(\frac{\rho_{S}}{|q|}\right)q^{i}.
\end{equation}
Comparing this expression with $v^{i}=C_{\mathrm{geom}}q^{i}$, we
obtain

\begin{equation}
C_{\mathrm{geom}}=\frac{\rho_{S}}{|q|}=\frac{2\pi r_{h}\rho_{S}}{\mathcal{N}}=\frac{S_{\mathrm{BH}}^{\left(0\right)}}{\mathcal{N}}.
\end{equation}
Equivalently,

\begin{equation}
\rho_{S}\hat{q}^{i}=C_{\mathrm{geom}}q^{i}.
\end{equation}
The two expressions therefore do not represent different correspondences.
Rather, they are two normalizations of the same entropy-flow field.
The form

\begin{equation}
v_{\mathrm{phys}}^{i}=\rho_{S}\hat{q}^{i},
\end{equation}
separates the direction selected by the stationary F1 dynamics from
the entropy capacity $\rho_{S}$ given by the off-shell string construction.
It thus makes the local geometric structure of the flow manifest.
In the present three-dimensional problem,

\begin{equation}
dS_{\max}=\rho_{S}dL_{S},\qquad\rho_{S}=\frac{e^{-2\phi_{0}}}{4G_{N}^{\left(3\right)}},
\end{equation}
so that $\rho_{S}$ specifies the maximal entropy $S_{\max}$ that
can cross a unit element $L_{S}$ of string-frame length, whereas
$\widehat{q}^{i}$ specifies the unit direction of the string flow.

On the other hand, the form

\begin{equation}
v_{\mathrm{phys}}^{i}=C_{\mathrm{geom}}q^{i},
\end{equation}
makes explicit how the unnormalized string-charge current is converted
into an entropy flow. If every string carries one unit of Kalb--Ramond
charge, then

\begin{equation}
C_{\mathrm{geom}}=\frac{S_{\mathrm{BH}}^{\left(0\right)}}{\mathcal{N}},
\end{equation}
can be interpreted as the entropy capacity allocated per unit of string
charge. Under the assumptions of unit charge, common orientation,
and uniform coarse graining, this is also the entropy capacity assigned
to each string. More generally, $C_{\mathrm{geom}}$ is the entropy
weight per unit of normalized string charge.

This relation admits a direct coarse-grained geometric interpretation.
For $\mathcal{N}$ equally weighted unit-charge strings, the string-frame
horizon circumference is

\begin{equation}
A_{H}^{\left(S\right)}=2\pi r_{h},
\end{equation}
and the portion assigned to each string is
\begin{equation}
\Delta L_{H}=\frac{A_{H}^{\left(S\right)}}{\mathcal{N}}=\frac{2\pi r_{h}}{\mathcal{N}},
\end{equation}
Since the entropy capacity per unit string-frame length is $\rho_{S}$,
the entropy capacity associated with each unit-charge string is

\begin{equation}
C_{\mathrm{geom}}=\rho_{S}\Delta L_{H}=\rho_{S}\frac{2\pi r_{h}}{\mathcal{N}}=\frac{S_{\mathrm{BH}}^{\left(0\right)}}{\mathcal{N}}.
\end{equation}
The relation between the two normalizations may therefore be summarized
as

\begin{equation}
\underbrace{\rho_{S}}_{\substack{\text{entropy capacity per unit}\\
\text{string-frame length}
}
}\times\underbrace{\frac{2\pi r_{h}}{\mathcal{N}}}_{\substack{\text{horizon length assigned}\\
\text{to each string}
}
}=\underbrace{C_{\mathrm{geom}}}_{\substack{\text{entropy capacity per unit}\\
\text{string charge}
}
}.
\end{equation}
In other words, $C_{\mathrm{geom}}$ is the product of the local entropy
capacity per unit string-frame length and the share of the horizon
circumference allocated to each unit of string charge.

As a consistency check, the total flux of the resulting entropy flow
is 

\begin{equation}
\Phi\left[v_{\mathrm{phys}}\right]=C_{\mathrm{geom}}\Phi\left[q\right]=\frac{\rho_{S}}{|q|}\mathcal{N}=2\pi r_{h}\rho_{S}=S_{\mathrm{BH}}^{\left(0\right)}.
\end{equation}
The central result is therefore that, for the stationary BTZ string
ensemble, $C_{\mathrm{geom}}$ is no longer an independent global
normalization. It factorizes into the local entropy capacity $\rho_{S}$,
determined by the off-shell string construction, and the inverse norm
$1/|q|$, determined by the stationary F1 dynamics: 
\begin{equation}
C_{\mathrm{geom}}=\frac{\rho_{S}}{|q|}.
\end{equation}
The global relation $C_{\mathrm{geom}}\mathcal{N}=S_{\mathrm{BH}}^{\left(0\right)}$
then follows directly from these two local ingredients.

\section{One more example: BTZ with mixed flux}

\label{subsec:extflow-mixed} \label{sec:extflow}

We now extend the construction of Section~\ref{sec:physical-string-BTZ}
to BTZ supported by both NS--NS and Ramond--Ramond (RR) three-form
fluxes. We keep the string-frame geometry and the dilaton fixed and
vary the fraction of NS--NS flux. This separates the effect of the
background on the F1 embedding from the geometric entropy capacity. 

\subsection{The mixed-flux action and the BTZ background}

The mixed-flux solution is a type IIB background. The fields include
the string-frame metric, the dilaton, the Kalb--Ramond field $B$,
and the RR two-form $C^{\left(2\right)}$. In the sector with vanishing
axion and gauge invariant five-form, the relevant two-derivative terms
of the bulk action are
\begin{equation}
\begin{split}S_{\mathrm{IIB}}^{\left(0\right)}=\frac{1}{16\pi G_{N}^{\left(10\right)}}\int d^{10}x\sqrt{-g_{10}}\bigg[e^{-2\phi}\left(R_{10}+4\left(\nabla\phi\right)^{2}-\frac{1}{12}H_{MNP}H^{MNP}\right)-\frac{1}{12}F_{MNP}^{\left(3\right)}F^{\left(3\right)MNP}\bigg],\end{split}
\label{eq:mixed action}
\end{equation}
where $H=dB$ and $F^{\left(3\right)}=dC^{\left(2\right)}$ in this
sector. The BTZ quotient of the mixed-flux $\mathrm{AdS}_{3}\times S^{3}\times T^{4}$
solution of this action is given by \cite{Cagnazzo:2012se,Hoare:2013pma}
\begin{equation}
dS_{10}^{2}=ds_{\mathrm{BTZ}}^{2}+\ell^{2}d\Omega_{3}^{2}+ds_{T^{4}}^{2},\qquad\phi=\phi_{0},\label{eq:extflow-mixed-product}
\end{equation}
where 
\begin{equation}
dS_{\mathrm{BTZ}}^{2}=-f\left(r\right)dt^{2}+\frac{dr^{2}}{f\left(r\right)}+r^{2}d\theta^{2},\qquad f\left(r\right)=\frac{r^{2}-r_{h}^{2}}{\ell^{2}},\qquad\theta\sim\theta+2\pi.\label{eq:extflow-mixed-background}
\end{equation}
Let $\varepsilon_{\mathrm{BTZ}}$ and $\varepsilon_{S^{3}}$ denote
the volume three-forms of the physical metrics, both with curvature
radius $\ell$. In particular, $\varepsilon_{\mathrm{BTZ}}=rdt\wedge dr\wedge d\theta$.
Choosing the flux orientation to agree with the previous section gives
\begin{align}
H & =\frac{2\eta}{\ell}\left(\varepsilon_{\mathrm{BTZ}}+\varepsilon_{S^{3}}\right),\nonumber \\
F^{\left(3\right)} & =\frac{2e^{-\phi_{0}}\sqrt{1-\eta^{2}}}{\ell}\left(\varepsilon_{\mathrm{BTZ}}+\varepsilon_{S^{3}}\right),\qquad0\leq\eta\leq1.\label{eq:extflow-mixed-fluxes}
\end{align}
The parameter $\eta$ denotes the NS--NS fraction and is distinct
from the projected current $q^{i}$. The two physical volume forms
have squared contractions $-3!$ and $+3!$, respectively. Consequently,
\begin{equation}
H^{2}=\left(F^{\left(3\right)}\right)^{2}=H\mathbin{\cdot}F^{\left(3\right)}=0,\qquad H\wedge F^{\left(3\right)}=0.
\end{equation}
The last two identities ensure that the axion and five-form equations
are satisfied when those fields are set to zero. The Bianchi identities
and the three-form equations hold because the physical volume forms
are closed and covariantly constant.

On this solution, the dilaton is constant and both three-form scalar
contractions vanish. The metric equation therefore reduces to
\begin{equation}
R_{MN}=\frac{1}{4}H_{MPQ}H_{N}{}^{PQ}+\frac{e^{2\phi_{0}}}{4}F_{MPQ}^{\left(3\right)}F_{N}^{\left(3\right)\,PQ}.\label{eq:extflow-mixed-Einstein}
\end{equation}
For BTZ indices $\mathrm{a},\mathrm{b}$ this gives 
\begin{equation}
R_{\mathrm{ab}}=-\frac{2}{\ell^{2}}\left[\eta^{2}+\left(1-\eta^{2}\right)\right]g_{\mathrm{ab}}=-\frac{2}{\ell^{2}}g_{\mathrm{ab}}.\label{eq:extflow-mixed-support}
\end{equation}
The sphere equation has the opposite curvature sign, and the torus
is flat. The dilaton equation is satisfied since $R_{10}=H^{2}=0$.
Thus the metric is unchanged as $\eta$ varies.

Integrating over the fixed internal geometry gives the following restricted
three-dimensional action:
\begin{equation}
\begin{split}S_{\mathrm{3D}}^{\left(0\right)}=\frac{1}{16\pi G_{N}^{\left(3\right)}}\int d^{3}x\sqrt{-g}\bigg\{ e^{-2\phi}\left[R+4\left(\nabla\phi\right)^{2}-\frac{1}{12}H_{\mathrm{abc}}H^{\mathrm{abc}}+\frac{6-2\eta^{2}}{\ell^{2}}\right]\\
-\frac{1}{12}F_{\mathrm{abc}}^{\left(3\right)}F^{\left(3\right)\,\mathrm{abc}}-\frac{2e^{-2\phi_{0}}\left(1-\eta^{2}\right)}{\ell^{2}}\bigg\},\qquad\frac{1}{G_{N}^{(3)}}=\frac{V_{\mathrm{int}}^{(S)}}{G_{N}^{(10)}},
\end{split}
\label{eq:extflow-mixed-3d-action}
\end{equation}
where $V_{\mathrm{int}}^{(S)}\equiv V_{S^{3}}V_{T^{4}}$. The sphere
curvature gives $6/\ell^{2}$, while the frozen internal NS--NS and
RR fluxes give the remaining potential terms. In this restricted action,
$\phi\left(x\right)$ is varied while $\phi_{0}$, the internal metric,
and the internal fluxes are held at their background values. At $\eta=1$
it reduces to the pure NS--NS potential $4/\ell^{2}$.

The BTZ metric together with $\phi=\phi_{0}$ solves these external
equations with potentials
\begin{align}
B\big|_{\mathrm{BTZ}} & =-b_{\eta}\left(r\right)dt\wedge d\theta, & b_{\eta}\left(r\right) & =\frac{\eta\left(r^{2}-r_{h}^{2}\right)}{\ell},\nonumber \\
C^{\left(2\right)}\big|_{\mathrm{BTZ}} & =-c_{\eta}\left(r\right)dt\wedge d\theta, & c_{\eta}\left(r\right) & =\frac{e^{-\phi_{0}}\sqrt{1-\eta^{2}}\left(r^{2}-r_{h}^{2}\right)}{\ell}.\label{eq:extflow-mixed-potentials}
\end{align}
Both potentials vanish at the horizon and reproduce the external parts
of Eq. (\ref{eq:extflow-mixed-fluxes}). Their internal components
produce the sphere fluxes and have zero pullback to the worldsheets
with fixed internal coordinates.

For clarity, the scalar contractions in three dimensions are nonzero:
\begin{equation}
H_{\mathrm{abc}}H^{\mathrm{abc}}=-\frac{24\eta^{2}}{\ell^{2}},\qquad\left(F_{\mathrm{ext}}^{\left(3\right)}\right)^{2}=-\frac{24e^{-2\phi_{0}}\left(1-\eta^{2}\right)}{\ell^{2}}.
\end{equation}
Writing $U_{\eta}=\left(6-2\eta^{2}\right)/\ell^{2}$, the dilaton
equation at $\phi=\phi_{0}$ is 
\begin{equation}
R-\frac{1}{12}H_{\mathrm{ext}}^{2}+U_{\eta}=\frac{-6+2\eta^{2}+6-2\eta^{2}}{\ell^{2}}=0.
\end{equation}
If $V_{\mathrm{RR}}=2e^{-2\phi_{0}}\left(1-\eta^{2}\right)/\ell^{2}$
denotes the last potential in Eq. (\ref{eq:extflow-mixed-3d-action}),
the constant-dilaton metric equation, after using the dilaton equation,
is 
\begin{align}
R_{\mathrm{ab}}= & \frac{1}{4}H_{\mathrm{acd}}H_{\mathrm{b}}{}^{\mathrm{cd}}+\frac{e^{2\phi_{0}}}{4}F_{\mathrm{acd}}^{\left(3\right)}F_{\mathrm{b}}^{\left(3\right)}{}^{\mathrm{cd}}\nonumber \\
 & -\frac{e^{2\phi_{0}}}{24}g_{\mathrm{ab}}\left(F_{\mathrm{ext}}^{\left(3\right)}\right)^{2}-\frac{e^{2\phi_{0}}}{2}V_{\mathrm{RR}}g_{\mathrm{ab}}=-\frac{2}{\ell^{2}}g_{\mathrm{ab}}.
\end{align}
The two RR trace terms cancel on the background. This verifies the
BTZ solution directly in the restricted three-dimensional action.
The continuous family is understood classically. Its flux quantization
restricts the allowed values in a quantum compactification.

\subsection{The F1 source action and its regular stationary solution}

We add fundamental string sources to the ten-dimensional type IIB
action (\ref{eq:mixed action}): 
\begin{equation}
S_{\mathrm{tot}}^{\left(0\right)}=S_{\mathrm{IIB}}^{\left(0\right)}+S_{\sigma},\qquad S_{\sigma}=\sum_{k=1}^{\mathcal{N}}S_{\mathrm{F1}}\left[X_{k}\right],\label{eq:extflow-mixed-total-action}
\end{equation}
with 

\begin{equation}
S_{\mathrm{F1}}=-\frac{1}{4\pi\alpha^{\prime}}\int d^{2}\sigma\left(\sqrt{-\gamma}\gamma^{mn}\partial_{m}X^{\mathrm{a}}\partial_{n}X^{\mathrm{b}}g_{\mathrm{ab}}+\epsilon^{mn}\partial_{m}X^{\mathrm{a}}\partial_{n}X^{\mathrm{b}}B_{\mathrm{ab}}\right),\qquad\epsilon^{\tau\sigma}=+1,\label{eq:mixed source}
\end{equation}
where the worldsheet fermions are set to zero. RR fields enter the
full superstring action through fermionic couplings, so the F1 action
has no bosonic electric coupling to $C^{\left(2\right)}$ \cite{Hoare:2013pma}.
In other words, the RR field supports the background, while $B$ supplies
the flux in the embedding equation (\ref{eq:mixed source}). Moreover,
we also use the probe limit in our calculations. The embeddings solve
the F1 equations in this background, and the response of the background
to $S_{\sigma}$ is neglected.

Choose static gauge separately on each exterior worldsheet, 
\begin{equation}
X^{t}=\tau,\qquad X^{r}=\sigma=r,\qquad X^{\theta}=\theta_{\eta}\left(r\right),\qquad X^{\mathrm{internal}}=\mathrm{constant}.\label{eq:extflow-mixed-static-gauge}
\end{equation}
Writing the induced worldsheet metric as $\gamma_{mn}^{\mathrm{ind}}=g_{\mathrm{ab}}\partial_{m}X^{\mathrm{a}}\partial_{n}X^{\mathrm{b}}$,
we obtain 
\begin{equation}
\gamma_{\tau\tau}^{\mathrm{ind}}=-f,\qquad\gamma_{\sigma\sigma}^{\mathrm{ind}}=f^{-1}+r^{2}\theta_{\eta}^{\prime2},\qquad\gamma_{\tau\sigma}^{\mathrm{ind}}=0,\qquad-\det\gamma^{\mathrm{ind}}=1+fr^{2}\theta_{\eta}^{\prime2}.\label{eq:extflow-mixed-induced}
\end{equation}
Eliminating the auxiliary metric in Eq. (\ref{eq:mixed source}) gives
\begin{equation}
S_{\mathrm{F1}}=-\frac{1}{2\pi\alpha^{\prime}}\int dtdr\left(\sqrt{1+fr^{2}\theta_{\eta}^{\prime2}}-b_{\eta}\theta_{\eta}^{\prime}\right).\label{eq:extflow-mixed-reduced-source}
\end{equation}
The sign of the second term follows from $B=-b_{\eta}dt\wedge d\theta$
and the chosen worldsheet orientation. 

Since $\theta_{\eta}$ is cyclic, its equation has the first integral
\begin{equation}
\frac{fr^{2}\theta_{\eta}^{\prime}}{\sqrt{1+fr^{2}\theta_{\eta}^{\prime2}}}-b_{\eta}=k_{\theta}.\label{eq:extflow-mixed-first-integral}
\end{equation}
Squaring this equation gives 
\begin{equation}
\theta_{\eta}^{\prime2}=\frac{\left(b_{\eta}+k_{\theta}\right)^{2}}{fr^{2}\left[fr^{2}-\left(b_{\eta}+k_{\theta}\right)^{2}\right]}.\label{eq:extflow-mixed-squared-integral}
\end{equation}
A branch reaching the horizon with finite $\theta_{\eta}^{\prime}$
has $k_{\theta}=0$, since $f\left(r_{h}\right)=b_{\eta}\left(r_{h}\right)=0$. 

Define the positive effective width 
\begin{equation}
\mathcal{D}_{\eta}^{2}\left(r\right)\equiv r^{2}-\frac{b_{\eta}^{2}}{f}=\left(1-\eta^{2}\right)r^{2}+\eta^{2}r_{h}^{2},\label{eq:extflow-width}
\end{equation}
and solving Eq. (\ref{eq:extflow-mixed-first-integral}) on the regular
branch gives 
\begin{equation}
\theta_{\eta}^{\prime}=\frac{b_{\eta}}{fr\mathcal{D}_{\eta}}=\frac{\eta\ell}{r\mathcal{D}_{\eta}},\qquad\sqrt{1+fr^{2}\theta_{\eta}^{\prime2}}=\frac{r}{\mathcal{D}_{\eta}}.\label{eq:extflow-mixed-direction}
\end{equation}
In particular, $\mathcal{D}_{\eta}\left(r_{h}\right)=r_{h}$, $\theta_{\eta}^{\prime}\left(r_{h}\right)=\eta\ell/r_{h}^{2}$,
and $\sqrt{-\det\gamma^{\mathrm{ind}}}\big|_{r_{h}}=1$. The apparent
singularity of static coordinates is removable. In ingoing coordinates
$v=t+\int dr/f$, the gauge $X^{v}=\tau$, $X^{r}=\sigma$ gives 
\begin{equation}
ds_{\mathrm{ws}}^{2}=-fd\tau^{2}+2d\tau dr+r^{2}\theta_{\eta}^{\prime2}dr^{2},\qquad\det\gamma^{\mathrm{ind}}=-\left(1+fr^{2}\theta_{\eta}^{\prime2}\right),\label{eq:extflow-mixed-horizon-regularity}
\end{equation}
which is nondegenerate at the future horizon. The pullback of the
potential is regular as well. These are on-shell probe F1 sources,
with the remaining embedding equations following from the worldsheet
reparametrization identities once the transverse equation and auxiliary-metric
equation hold.

\begin{figure}[h]
\begin{centering}
\includegraphics[scale=0.38]{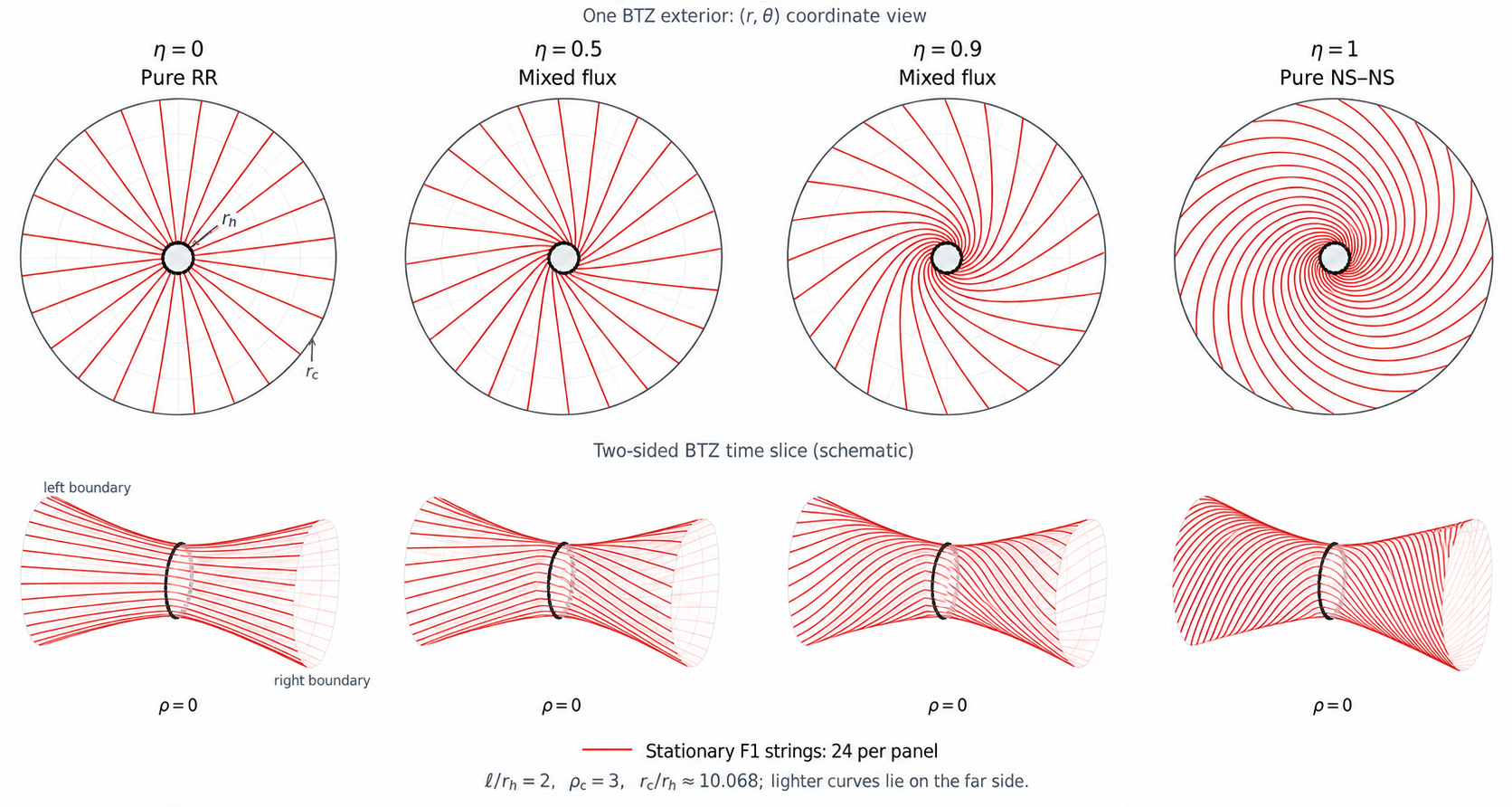}
\par\end{centering}
\caption{\label{fig:new source-1}Red solid curves show stationary F1 strings
for $\eta=0,0.5,0.9$ and $1$, from left to right. The strings interpolate
from radial curves to logarithmic spirals. The upper row shows one
BTZ exterior, and the lower row shows the two-sided spatial slice
schematically. }
\end{figure}

Integrating with $\theta_{\eta}\left(r_{h}\right)=\theta_{h}$ yields
\begin{equation}
\theta_{\eta}\left(r\right)=\theta_{h}+\frac{\ell}{r_{h}}\log\frac{\left(1+\eta\right)r}{\mathcal{D}_{\eta}\left(r\right)+\eta r_{h}}.\label{eq:extflow-mixed-profile}
\end{equation}
The limiting profiles $\eta=0,1$ are 
\begin{equation}
\theta_{0}\left(r\right)=\theta_{h},\qquad\theta_{1}\left(r\right)=\theta_{h}+\frac{\ell}{r_{h}}\log\frac{r}{r_{h}}.\label{eq:extflow-mixed-limits}
\end{equation}
Thus a radial F1 is an on-shell probe in the pure RR background. It
still carries Kalb--Ramond charge, defined by varying its source
coupling to $B$, even though the background value of $H$ vanishes.
For every $\eta>0$, a radial embedding would instead require $b_{\eta}^{\prime}=2\eta r/\ell=0$
and fails the stationary F1 equation. The physical F1 sources with
different choices of $\eta$ on a constant time slice of one BTZ exterior
and two-sided BTZ geometry is illustrated in fig. (\ref{fig:new source-1}).

For $0\leq\eta<1$ the total unwrapped angular sweep is finite: 
\begin{equation}
\theta_{\eta}\left(\infty\right)-\theta_{h}=\frac{\ell}{r_{h}}\mathrm{arc}\tanh\eta.\label{eq:extflow-mixed-sweep}
\end{equation}
It diverges as $\eta\to1$, where the logarithmic spiral is recovered.
As an additional check, the on-shell reduced integrand is 
\begin{equation}
\sqrt{1+fr^{2}\theta_{\eta}^{\prime2}}-b_{\eta}\theta_{\eta}^{\prime}=\frac{\mathcal{D}_{\eta}}{r}.\label{eq:extflow-mixed-onshell-action}
\end{equation}

\subsection{The physical string current, spatial projection and black hole entropy}

Now, we calculate the physical string current and its spatial projection
based on this solution. The source variation and the corresponding
current are 

\begin{equation}
j_{\eta}^{\mathrm{ab}}\left(x\right)=\frac{1}{\sqrt{-g\left(x\right)}}\sum_{k=1}^{\mathcal{N}}\int d\tau d\sigma\epsilon^{mn}\partial_{m}X_{k}^{\mathrm{a}}\partial_{n}X_{k}^{\mathrm{b}}\delta^{\left(3\right)}\left(x-X_{k}\right).
\end{equation}
These are the three-dimensional currents obtained after integrating
the ten-dimensional distribution over the internal coordinates. Take
$\theta_{h,k}$ uniformly distributed around the horizon circle. Every
profile is an angular translate of Eq. (\ref{eq:extflow-mixed-profile}),
so the continuum replacement 
\begin{equation}
\sum_{k=1}^{\mathcal{N}}\delta_{2\pi}\left(\theta-\theta_{\eta,k}\left(r\right)\right)\longrightarrow\frac{\mathcal{N}}{2\pi}.\label{eq:extflow-mixed-smearing}
\end{equation}
Using $\sqrt{-g}=r$ and the static gauge, the nonzero independent
components become 
\begin{equation}
j_{\eta}^{0r}=\frac{\mathcal{N}}{2\pi r},\qquad j_{\eta}^{0\theta}=\frac{\mathcal{N}}{2\pi r}\theta_{\eta}^{\prime},\qquad j_{\eta}^{r\theta}=0.\label{eq:extflow-mixed-current-components}
\end{equation}
The angular component is therefore part of the physical current, just
as in the pure NS--NS calculation. On the static slice, 
\begin{equation}
h_{ij}dx^{i}dx^{j}=\frac{dr^{2}}{f}+r^{2}d\theta^{2},\qquad\sqrt{h}=\frac{r}{\sqrt{f}},\qquad N_{\mathrm{ADM}}=\sqrt{f}.\label{eq:extflow-mixed-slice}
\end{equation}
If $n_{\mu}$ is its future-directed unit normal, the same projection
$q_{\eta}^{\mu}=-n_{\nu}j_{\eta}^{\nu\mu}$ gives 
\begin{equation}
q_{\eta}^{r}=N_{\mathrm{ADM}}j_{\eta}^{0r}=\frac{\mathcal{N}\sqrt{f}}{2\pi r},\qquad q_{\eta}^{\theta}=N_{\mathrm{ADM}}j_{\eta}^{0\theta}=\frac{\mathcal{N}\eta\ell\sqrt{f}}{2\pi r^{2}\mathcal{D}_{\eta}}=\theta_{\eta}^{\prime}q_{\eta}^{r}.\label{eq:extflow-mixed-projected-current}
\end{equation}
This current is conserved in the exterior, away from any string endpoints:
\begin{equation}
\nabla_{i}q_{\eta}^{i}=\frac{1}{\sqrt{h}}\left[\partial_{r}\left(\sqrt{h}q_{\eta}^{r}\right)+\partial_{\theta}\left(\sqrt{h}q_{\eta}^{\theta}\right)\right]=\frac{\sqrt{f}}{r}\partial_{r}\left(\frac{\mathcal{N}}{2\pi}\right)=0.\label{eq:extflow-mixed-conservation}
\end{equation}
For a circle $C_{r}$, take its normal toward increasing $r$, $n^{i}=\sqrt{f}\left(\partial_{r}\right)^{i}$,
and $dA_{S}=rd\theta$. Then 
\begin{equation}
\Phi\left[q_{\eta}\right]=\int_{0}^{2\pi}\frac{q_{\eta}^{r}}{\sqrt{f}}rd\theta=\mathcal{N}.\label{eq:extflow-mixed-charge-flux}
\end{equation}
The full current norm follows directly from the embedding equation:
\begin{equation}
\begin{split}|q_{\eta}|^{2} & =\frac{\left(q_{\eta}^{r}\right)^{2}}{f}+r^{2}\left(q_{\eta}^{\theta}\right)^{2}\\
 & =\left(\frac{\mathcal{N}}{2\pi r}\right)^{2}\left(1+fr^{2}\theta_{\eta}^{\prime2}\right)=\left(\frac{\mathcal{N}}{2\pi\mathcal{D}_{\eta}}\right)^{2}.
\end{split}
\label{eq:extflow-mixed-current-norm}
\end{equation}
Moreover, it is worth noting that its horizon value is independent
of $\eta$, but its exterior value is constant only for pure NS--NS
flux: 
\begin{equation}
|q_{\eta}|_{h}=\frac{\mathcal{N}}{2\pi r_{h}},\qquad|q_{0}|=\frac{\mathcal{N}}{2\pi r},\qquad|q_{1}|=\frac{\mathcal{N}}{2\pi r_{h}}.\label{eq:extflow-mixed-current-limits}
\end{equation}
Unlike the pure NS--NS result of Section (\ref{sec:physical-string-BTZ}),
the pointwise unit normalization of the mixed-flux current fails to
preserve conservation. For $\widehat{q}_{\eta}^{i}=q_{\eta}^{i}/|q_{\eta}|$,
\begin{equation}
\nabla_{i}\left(\rho_{S}\widehat{q}_{\eta}^{i}\right)=\frac{\rho_{S}\sqrt{f}}{r}\mathcal{D}_{\eta}^{\prime}=\frac{\rho_{S}\left(1-\eta^{2}\right)\sqrt{f}}{\mathcal{D}_{\eta}},\label{eq:extflow-mixed-unit-divergence}
\end{equation}
which is nonzero in the exterior when $\eta<1$. Therefore, we need
to use a constant conversion coefficient along the strings as an alternative:
\begin{equation}
v_{\mathrm{phys},\eta}^{i}=C_{\mathrm{geom}}q_{\eta}^{i}.\label{eq:extflow-mixed-correspondence}
\end{equation}
In general, conservation requires $q_{\eta}^{i}\partial_{i}C_{\mathrm{geom}}=0$,
so the coefficient is constant along each string. Within the axisymmetric
uniform ensemble, $C_{\mathrm{geom}}=C_{\mathrm{geom}}\left(r\right)$,
and this reduces to $q_{\eta}^{r}\partial_{r}C_{\mathrm{geom}}=0$,
so it is constant on the connected exterior. Within the uniform ensemble,
the largest positive constant compatible with the independently derived
capacity is determined by 

\begin{eqnarray}
C_{\mathrm{geom}}^{\max} & = & \inf_{r\geq r_{h}}\frac{\rho_{S}}{|q_{\eta}\left(r\right)|}=\frac{2\pi\rho_{S}}{\mathcal{N}}\inf_{r\geq r_{h}}\mathcal{D}_{\eta}\left(r\right)\nonumber \\
 & = & \frac{2\pi r_{h}\rho_{S}}{\mathcal{N}}=\frac{\rho_{S}}{|q_{\eta}|_{h}},
\end{eqnarray}
where we use 
\begin{equation}
\mathcal{D}_{\eta}^{2}-r_{h}^{2}=\left(1-\eta^{2}\right)\left(r^{2}-r_{h}^{2}\right)\geq0.\label{eq:extflow-mixed-criterion}
\end{equation}
Thus the coefficient follows from the local norm constraint and the
computed physical current before its total entropy flux is evaluated.
We henceforth write this maximal constant simply as $C_{\mathrm{geom}}$.

The resulting physical flow is 
\begin{equation}
v_{\mathrm{phys},\eta}=\rho_{S}\frac{r_{h}\sqrt{f}}{r}\left(\partial_{r}+\frac{\eta\ell}{r\mathcal{D}_{\eta}}\partial_{\theta}\right)=\rho_{S}\frac{r_{h}}{\mathcal{D}_{\eta}}\widehat{q}_{\eta},\label{eq:extflow-mixed-flow}
\end{equation}
and obeys 
\begin{equation}
\nabla_{i}v_{\mathrm{phys},\eta}^{i}=0,\qquad\frac{|v_{\mathrm{phys},\eta}|}{\rho_{S}}=\frac{r_{h}}{\mathcal{D}_{\eta}}\leq1.\label{eq:extflow-mixed-admissibility}
\end{equation}
For $\eta<1$, the inequality is strict at every $r>r_{h}$. For $\eta=1$,
it is saturated throughout the exterior. Consequently, maximal transport
does not require local saturation away from the entropy bottleneck.
The number $\mathcal{N}$ fixes the normalization of the charge current
and cancels from the entropy flow. 

We now compute the entropy flux directly, using the same circle normals
as in Eq. (\ref{eq:extflow-mixed-charge-flux}): 

\begin{eqnarray}
\Phi\left[v_{\mathrm{phys},\eta}\right] & = & \int_{0}^{2\pi}\frac{v_{\mathrm{phys},\eta}^{r}}{\sqrt{f}}rd\theta=2\pi r_{h}\rho_{S}\nonumber \\
 & = & \frac{e^{-2\phi_{0}}\left(2\pi r_{h}\right)}{4G_{N}^{\left(3\right)}}=S_{\mathrm{BH}}^{(0)}.
\end{eqnarray}
This equals the ten-dimensional entropy $e^{-2\phi_{0}}\left(2\pi r_{h}\right)V_{\mathrm{int}}^{\left(S\right)}/\left(4G_{N}^{\left(10\right)}\right)$.
The agreement holds for the thermal entropy, or equivalently the entanglement
entropy of one boundary of two-sided BTZ. 

The maximality of this flow follows from the same argument used for
pure NS--NS BTZ. The flow is conserved and satisfies the norm bound
throughout the exterior, 

\begin{equation}
\nabla_{i}v_{\mathrm{phys},\eta}^{i}=0,\qquad|v_{\mathrm{phys},\eta}|\leq\rho_{S}.
\end{equation}
Its orthonormal components are 
\begin{equation}
v_{\mathrm{phys},\eta}^{\hat{r}}=\rho_{S}\frac{r_{h}}{r},\qquad v_{\mathrm{phys},\eta}^{\hat{\theta}}=\frac{\rho_{S}\eta r_{h}\sqrt{r^{2}-r_{h}^{2}}}{r\mathcal{D}_{\eta}}.
\end{equation}
Taking the horizon limit shows that the flow is normal to the bifurcation
circle and saturates its capacity:
\begin{equation}
v_{\mathrm{phys},\eta}^{i}\big|_{\Gamma_{H}}=\rho_{S}n^{i}.
\end{equation}
 For any admissible flow $w^{i}$, conservation and the norm bound
therefore imply 

\begin{equation}
\Phi\left[w\right]\leq\int_{\Gamma_{H}}\rho_{S}dA_{S}=2\pi r_{h}\rho_{S}=\Phi\left[v_{\mathrm{phys},\eta}\right].
\end{equation}
Thus, the physical F1 current realizes a global maximal flow for every
$0\leq\eta\leq1$, with entropy flux $\Phi\left[v_{\mathrm{phys},\eta}\right]=S_{\mathrm{BH}}^{\left(0\right)}$.

Finally, we wish to note that the entropy flow has a smooth extension
across the two-sided spatial slice. Set $r=r_{h}\cosh\rho$, with
signed $\rho$, and define 
\begin{equation}
h_{ij}dx^{i}dx^{j}=\ell^{2}d\rho^{2}+r_{h}^{2}\cosh^{2}\rho d\theta^{2},\qquad W_{\eta}\left(\rho\right)=\frac{\mathcal{D}_{\eta}}{r_{h}}=\sqrt{1+\left(1-\eta^{2}\right)\sinh^{2}\rho}.\label{eq:extflow-two-sided-coordinates}
\end{equation}
For a flow oriented from the left exterior to the right, 
\begin{equation}
v_{\mathrm{phys},\eta}^{\rho}=\frac{\rho_{S}}{\ell\cosh\rho},\qquad v_{\mathrm{phys},\eta}^{\theta}=\frac{\eta\rho_{S}\sinh\rho}{r_{h}\cosh^{2}\rho W_{\eta}},\qquad|v_{\mathrm{phys},\eta}|=\frac{\rho_{S}}{W_{\eta}}.\label{eq:extflow-two-sided-flow}
\end{equation}
These are smooth functions at $\rho=0$, where the flow is normal
to the bifurcation circle. The outward left and right boundary fluxes
are $-S_{\mathrm{BH}}^{\left(0\right)}$ and $+S_{\mathrm{BH}}^{\left(0\right)}$,
where the opposite signs reflect the direction of the flow relative
to the outward boundary normals. The entropies of the two boundaries
are both positive and equal: $S_{L}=\Phi_{L}\left[-v_{\mathrm{phys},\eta}\right]=S_{\mathrm{BH}}^{\left(0\right)}$
and $S_{R}=\Phi_{R}\left[v_{\mathrm{phys},\eta}\right]=S_{\mathrm{BH}}^{\left(0\right)}$.
This is a continuation of the spatial flow and the static worldsheet
gauge is understood separately in the exterior charts.

The mixed-flux result highlights the role of physical string dynamics
in determining entropy-flow lines. Varying the NS--NS fraction $0\leq\eta\leq1$
changes the stationary F1 configurations, from radial strings in the
pure RR background to logarithmic spirals in the pure NS--NS background.
With the BTZ metric, dilaton, and local entropy capacity held fixed,
the corresponding conserved flows all saturate the horizon capacity
and carry the same maximal entropy flux, $2\pi r_{h}\rho_{S}=S_{\mathrm{BH}}^{\left(0\right)}$.
Different flux mixtures therefore select different physical realizations
of the same geometric entropy problem. This construction complements
the geometric bit-thread prescription by providing a dynamical origin
for a family of maximal flows.

\section{Discussion and conclusion}

\label{sec:conclusion}

In this paper, we have connected the local capacity of an entropy
flow to the off-shell closed string sphere functional and constructed
maximal-flow representatives from stationary fundamental-string sources.
At genus zero and two-derivative order, the conical variation of the
sphere-derived target space action gives 
\begin{equation}
S_{\mathrm{loc}}^{(0)}\left[m\right]=\int_{m}\rho_{S}dA_{S},\qquad\rho_{S}=\frac{e^{-2\phi}}{4G_{N}}.
\end{equation}
In a static holographic background, the semiclassical replica construction
selects the weighted minimal cut, and flow--cut duality turns its
entropy functional into the string-frame bound $|v|\leq\rho_{S}$.
This places the entropy capacity used in the string charge density/bit
threads correspondence on the same sphere-derived footing as the gravitational
entropy functional.

The NS--NS BTZ calculation gives the dynamical realization. A regular
stationary F1 forms a logarithmic spiral, and a uniform oriented ensemble
has the spatially constant current norm $|q|=\mathcal{N}/\left(2\pi r_{h}\right)$.
Its unit normalization remains conserved. Combining it with the constant
capacity gives 
\begin{equation}
v_{\mathrm{phys}}^{i}=\rho_{S}\widehat{q}^{i},\qquad C_{\mathrm{geom}}=\frac{\rho_{S}}{|q|},\qquad\Phi\left[v_{\mathrm{phys}}\right]=2\pi r_{h}\rho_{S}=S_{\mathrm{BH}}^{\left(0\right)}.
\end{equation}
The flow is normal to the horizon, saturates its capacity, and attains
the horizon upper bound on every admissible flow. Its flux is therefore
the dilaton-weighted BTZ entropy. Moreover, the conversion coefficient
$C_{\mathrm{geom}}$ follows from the independently determined capacity
and the stationary current.

We then extended the calculation to BTZ with mixed NS--NS/RR flux
in type IIB theory. We specify the bulk action, the complementary
fluxes supporting the fixed geometry, and the classical F1 source
action. The RR field supports the metric alongside the NS--NS field,
while the bosonic F1 embedding couples directly to the metric and
$B$. Solving the source equation in static gauge gives 
\begin{equation}
\theta_{\eta}^{\prime}=\frac{\eta\ell}{r\mathcal{D}_{\eta}},\qquad\mathcal{D}_{\eta}^{2}=\left(1-\eta^{2}\right)r^{2}+\eta^{2}r_{h}^{2},\qquad|q_{\eta}|=\frac{\mathcal{N}}{2\pi\mathcal{D}_{\eta}}.
\end{equation}
The regular profile interpolates from a physical radial F1 at pure
RR flux to a logarithmic spiral at pure NS--NS flux. For $\eta<1$
the current norm varies, so multiplying its unit direction by $\rho_{S}$
would violate conservation. The appropriate extension of the correspondence
is instead 
\begin{equation}
\begin{gathered}v_{\mathrm{phys},\eta}^{i}=\frac{\rho_{S}}{|q_{\eta}|_{h}}q_{\eta}^{i},\qquad\frac{|v_{\mathrm{phys},\eta}|}{\rho_{S}}=\frac{r_{h}}{\mathcal{D}_{\eta}}\leq1,\\
\Phi\left[v_{\mathrm{phys},\eta}\right]=\frac{e^{-2\phi_{0}}\left(2\pi r_{h}\right)}{4G_{N}^{\left(3\right)}}=S_{\mathrm{BH}}^{\left(0\right)}.
\end{gathered}
\end{equation}
The coefficient is the largest constant weight compatible with the
local capacity, determined from the physical current before its integrated
entropy flux is evaluated. The horizon upper bound, together with
the conserved admissible flow that saturates it, establishes global
maximality and horizon minimality. All flux fractions therefore give
the same entropy at fixed geometry, even though their physical strings
and the exterior norms of their flows differ. This distinguishes saturation
of the entropy bottleneck from saturation at every exterior point.

In other words, the mixed-flux extension exhibits a family of physical
realizations of the same BTZ entropy. Changing the NS--NS/RR mixture
changes the stationary F1 configurations, while the fixed metric and
dilaton keep the entropy capacity unchanged. For every flux fraction,
a constant normalization preserves current conservation, respects
the local norm bound, and saturates the horizon capacity. The resulting
maximal entropy flux is therefore independent of the flux mixture.
This result demonstrates how physical string dynamics determines the
flow lines while the geometric capacity fixes the maximal entropy
they carry.

The resulting physical picture is that the off-shell sphere functional
fixes the transverse entropy capacity, while the source worldsheets
provide dynamically determined lines carrying conserved string charge.
Assigning each line the capacity of its horizon share turns that charge
current into a maximal entropy flow. Physical strings thus furnish
explicit carriers of the flux representing entropy in the correspondence.
At the stated order, the stationary F1 ensembles constructed here
provide conserved currents that attain the entropy capacity fixed
by the sphere functional. Their geometric conversion coefficient is
fixed for the specified ensemble, while the microscopic entropy of
individual strings and the effects of backreaction, interactions,
higher genus, and higher derivatives remain outside this calculation.
Extending the explicit on-shell source construction to general boundary
subregions is a further problem beyond the complete-boundary BTZ examples
treated here.

\vspace{5mm}

\noindent {\bf Acknowledgements} 
HW is supported by NSFC Grant No.12105191. SY is supported by NSFC Grant No.12105031 and No.12547101.

\end{document}